\documentstyle[12pt,epsfig]{article}
\begin{document}
\begin{center}
\vspace*{0.5cm}

{\large\bf 
Determination of the pattern of nuclear binding from the
data on the lepton-nucleus deep inelastic scattering
}\\[0.7cm]

{\large G.I. Smirnov}$^{\mbox{1}}$\\[0.6cm]

Joint Institute for Nuclear Research, Dubna, Russia\\[0.8cm]
\end{center}

\begin{center}
\begin{abstract}

Nucleon structure function ratios $r^A(x)$ = $F_2^A(x)/F_2^{\rm D}(x)$
measured in the range of atomic masses $A \geq$ 4
are analyzed with the aim to determine the pattern of the $x$ and 
$A$ dependence of $F_2(x)$ modifications caused by nuclear environment.  
  It is found
that the $x$ and $A$ dependence of the deviations of the $r^A(x)$
from unity can be factorized in the entire range of $x$.
The characteristic feature  of the factorization is represented with  
the three cross-over points  $x_i$, $i$ = 1 -- 3 in which 
$r^A(x)$ = 1 independently of $A$.
In the range $x > 0.7$ the pattern of $r^A(x)$ is fixed with
$x_3$ = 0.84 $\pm$ 0.01.
  The pattern of the $x$ dependence is compared with theoretical
calculations of Burov, Molochkov and Smirnov to demonstrate that  
evolution of the nucleon structure as a function of $A$ occurs
in two steps, first for  $A \leq$ 4 and second for $A >$ 4. 
The long-standing problem of the origin of the EMC effect is understood
as the modification of the nucleon structure in the field responsible
for the binding forces in a three-nucleon system.
\end{abstract}

\vspace{2cm}
{\em Submitted to Eur. Phys. J. -- C}\\[0.5cm]
\end{center}

\vspace{5cm}
\rule {6cm}{0.5pt}

$^{\mbox{1}}${\em E-mail: G.Smirnov@cern.ch}
\newpage
  {\bf 1~ Introduction}\\

After nearly two decades of experimental and theoretical
investigations of the EMC effect, we have  rapidly accumulating
evidence that nuclear binding is the only physical mechanism 
which can be responsible for the modification of the nucleon
partonic structure by the nuclear medium. The modifications are
usually observed as a deviation from unity of the ratio
$r^{A/{\rm D}}(x) \equiv F_2^A(x)/F_2^{\rm D}(x)$,
where $F_2^A(x)$ and $F_2^{\rm D}(x)$ are the structure 
functions  per nucleon measured in a nucleus of mass $A$
and in a deuteron, respectively.

The publication \cite{gomez} of the results from SLAC added to the 
EMC effect controversy with the statement that the data on $r^A(x)$ 
does not directly  correlate with the binding energy per nucleon.
To clarify the role of binding forces I have suggested~\cite{sm95}
to determine the pattern of $r^{A/{\rm D}}(x)$, which clearly
conveys the message about saturation of modifications of $F_2(x)$
already at $A$ = 4. 
The saturation, according to~\cite{sm95}, had to
manifest itself not in the amplitude of the oscillations, but in the
pattern of the $x$ dependence of $r^{A/{\rm D}}(x)$, namely in the
 positions of the three cross-over points $x_i$, 
in which $r^{A/{\rm D}}(x_i)$ = 1. Such a pattern can be clearly seen
from the re-evaluated ratios $r^{A/{\rm D}}(x)$ of SLAC~\cite{gomez}
and NMC~\cite{ama95}.
 
In the present paper I analyze all the data on the ratio of $F_2^A$
and $F_2^{\rm D}$ structure functions available from
electron- and muon-nucleus deep inelastic scattering
 experiments (DIS) and extend my analysis to the range of $x \to 1$. 
Strictly speaking, the effect of modifications is a function of three
variables, $x$, $Q^2$ and $A$. I will use the data which belong to
the range 0.5~$< Q^2 <$~200 GeV$^2$ and are obtained on deuteron
and nuclear targets from $A$~=~4 to $A$~=~208.
Following the convention of the first EMC publication~\cite{aub83}
I disregard modifications of $F_2(x)$ in a deuterium nucleus.

As is known from experiments (see Refs.~\cite{gomez,ama95}), the pattern
of the EMC effect is $Q^2$ independent within a wide range of $x$.
This is consistent with the results of Ref.~\cite{kari}, in which
the $Q^2$ evolution of the modifications is considered in the leading
 order of QCD. It it shown in Ref.~\cite{kari}, that QCD evolution effect
in the ratio of tin-to-carbon structure functions is smaller than
experimental errors everywhere in the $x$ range, except for the
region of $x <$~0.05, in which the effect becomes comparable with
errors. This gives the arguments to investigate, below,
the  $x$ and $A$ dependence of nuclear effects after integrating them
 over $Q^2$.
The analysis includes recent measurements  of the ratios
$r^{A/{\rm C}}(x)$ $ \equiv F_2^A(x) / F_2^{\rm C}(x)$~\cite{arn96}.
 
  As a result, I determine the pattern of the modification of the nucleon
partonic structure  which evolves in $A$ independently of $x$ if $A >$ 4.
I also show that the missing patterns of the EMC effect 
in the lightest nuclei, which have been recently obtained
in Ref.~\cite{bms1}, are decisive for the understanding  the role
of nuclear binding both for $x$ and $A$ dependence of the effect as well
as for the understanding of the EMC effect origin.\\[0.5cm]

{\bf 2~ Distortion Pattern as a Function of $x$ and $A$}\\

As has been shown in Refs.~\cite{sm95,sm94}, the pattern of the
oscillations of $r^{A/{\rm D}}(x)$ has a universal shape in the range
 of 10$^{-3} < x < $ 0.7 and in the range of atomic masses $A \geq$ 4,
where the data from SLAC and NMC have been obtained.
Namely, the ratio  $F_2^A(x) / F_2^{\rm D}(x)$  
could be well approximated with the simplest phenomenological
 function,
\begin{equation}
r^{A/{\rm D}}(x) \equiv F_2^A(x) / F_2^{\rm D}(x) ~=~ 
x^{m_{sh}}(1+m_{anti})(1- m_{\mbox{\tiny EMC}} x), 
\label{x1x3}
\end{equation}
which contained only one free parameter $m_i$
for each of the three kinematic intervals, (1)~{\em nuclear shadowing},
(2)~{\em antishadowing} and (3)~{\em EMC effect}. By definition,
$m_i(A=2)$~=~0 and thus can serve for quantitative evaluation
of the $F_2(x)$  modifications in nuclei with $A >$~2.
%
\begin{center}
\begin{figure}[t]
\begin{center}
\begin{minipage}[t]{0.34 \linewidth}
\hspace*{-3cm}\mbox{\epsfysize=\hsize\epsffile{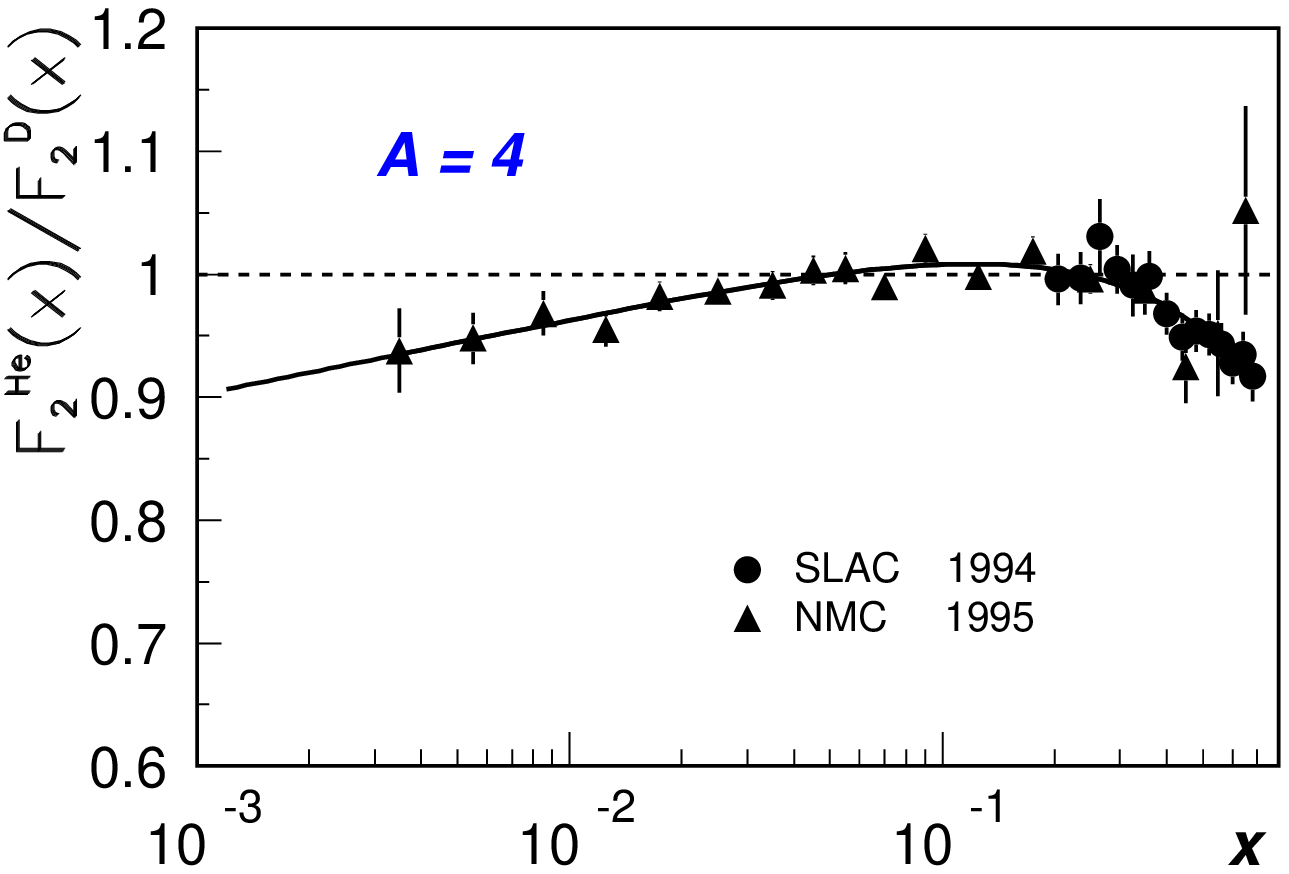}
\hspace{-0.5cm}\epsfysize=
\hsize\epsffile{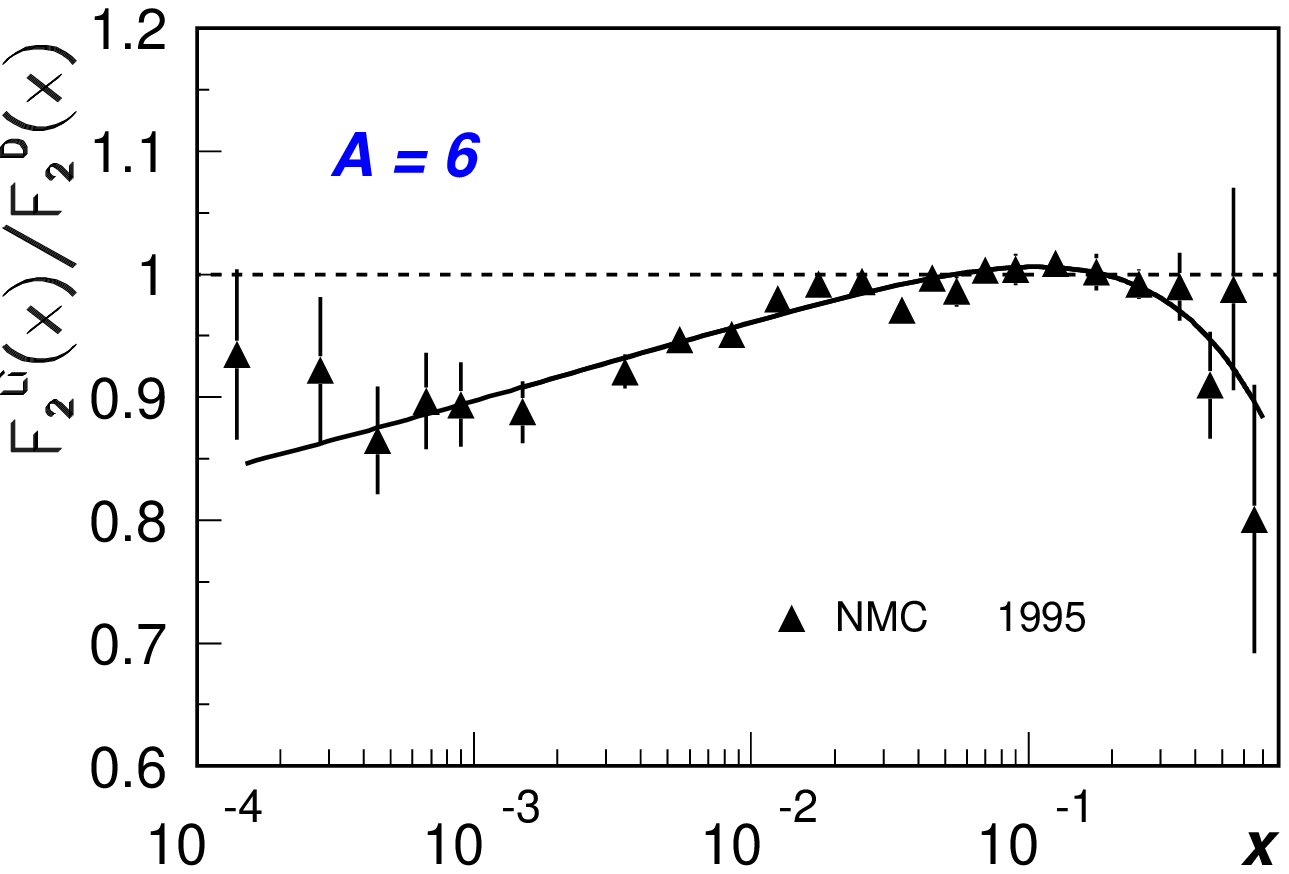}}
\end{minipage}
\end{center}
\vspace{-1.0cm}
\begin{center}
\begin{minipage}[t]{0.34   \linewidth}
\hspace*{-3cm}\mbox{\epsfysize=\hsize\epsffile{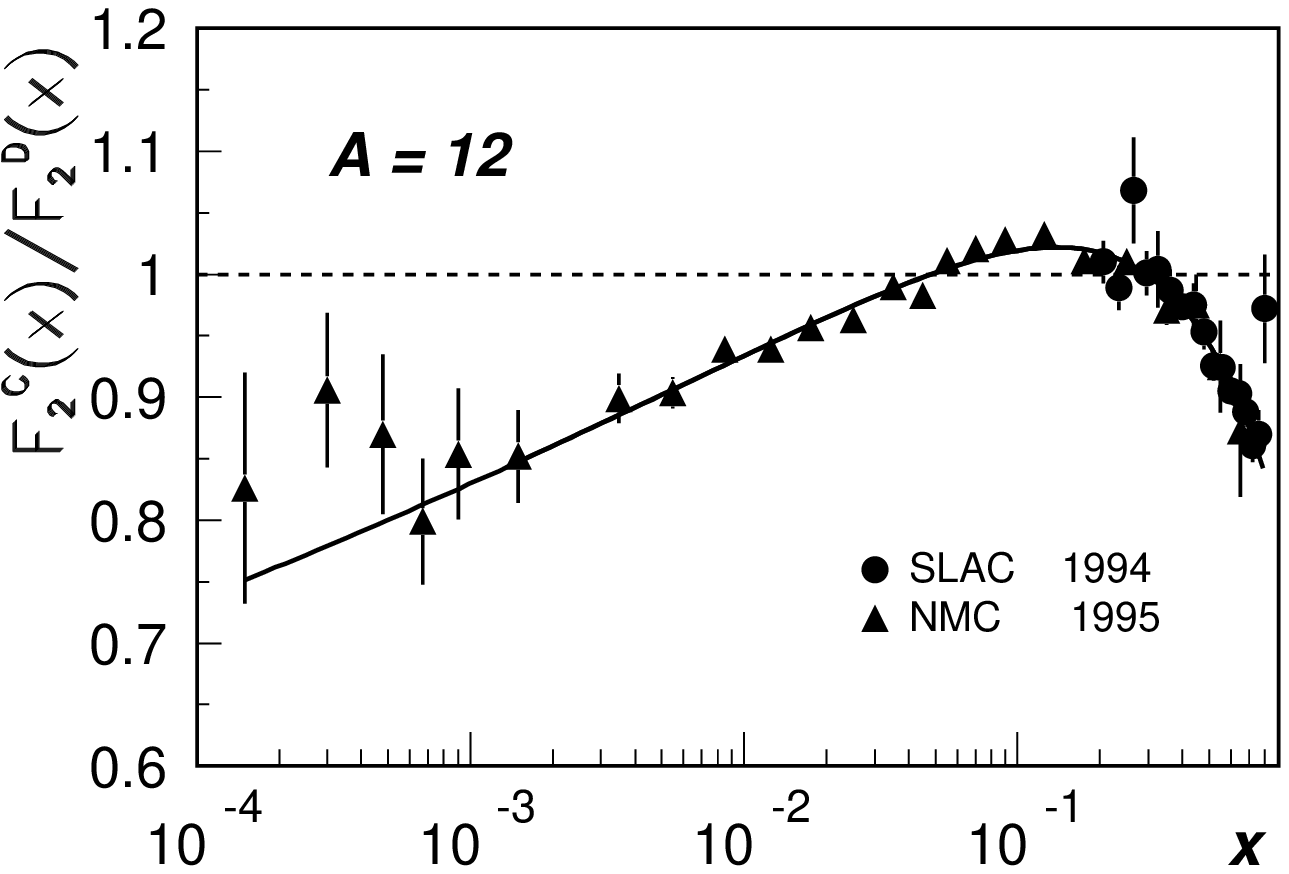}
\hspace{-0.5cm}\epsfysize=
\hsize\epsffile{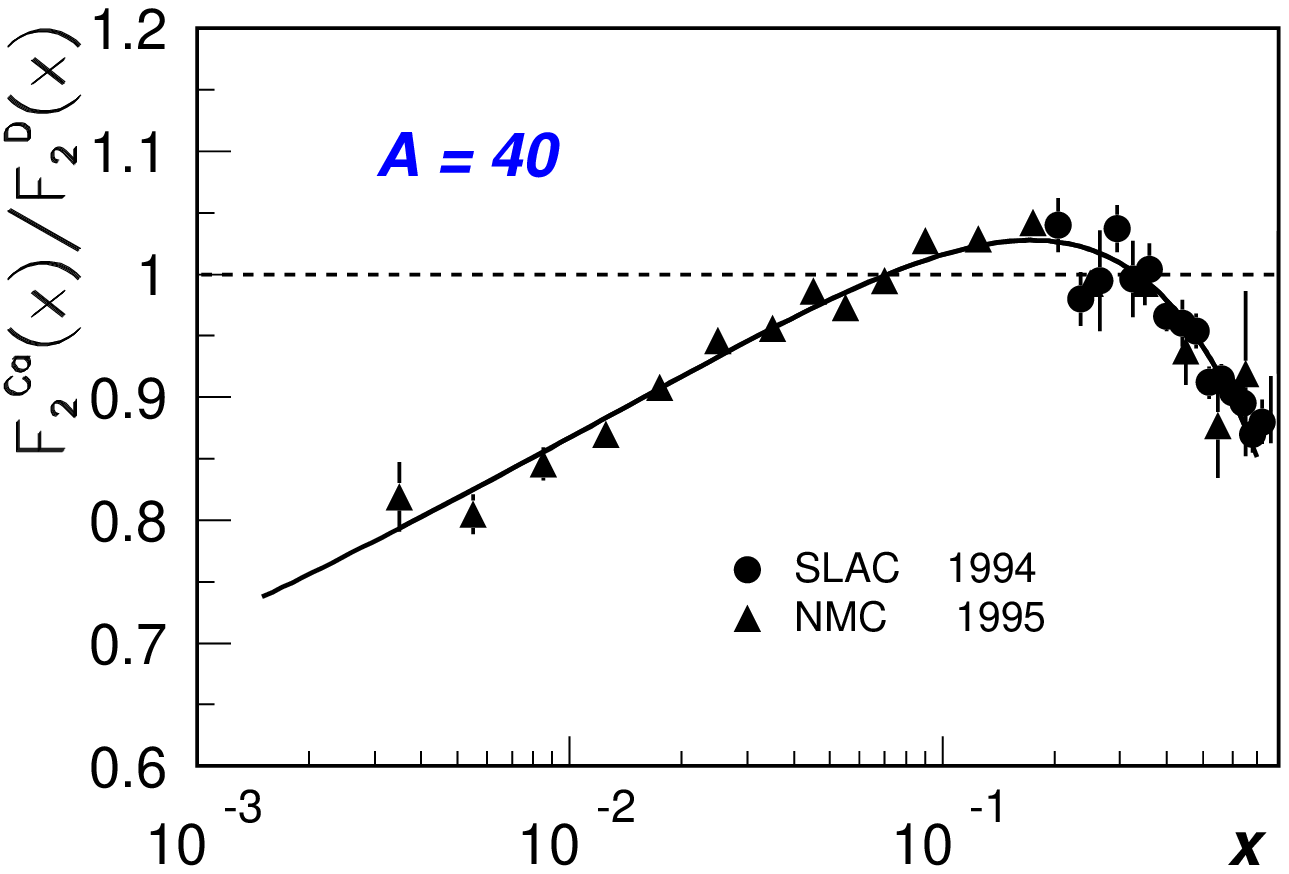}}
\end{minipage}
\end{center}
\vspace{-0.5cm}
\caption{ 
The results of the fit with Eq.~(1) of $F_2^A/F_2^{\rm D}$
measured by NMC and SLAC in the range 0.0001 $<x<$ 0.7 .}
\end{figure}
\end{center}

%
%
The agreement between the results of SLAC and NMC experiments 
significantly improved after NMC had presented the re-evaluated
data~\cite{ama95}. As a result, the agreement between the data and 
 Eq.~(\ref{x1x3})  (c.f. Figure 1)  has also improved, which allows for
better evaluation of parameters $m_i$ as a function of $A$.
Approximation of data for all available atomic masses $A$ with
Eq.~(\ref{x1x3}) turned out to be convenient for demonstration
of the factorization of $x$ and $A$ dependence of $F_2(x)$ 
modifications in nuclei in a wide range of $x$.
In other words, evolution of
the $x$ dependence of $r^A(x)$ ceased either at $A$ = 3
or $A$ = 4~\cite{sm95} which is very much consistent with the
phenomenon of {\em saturation} of nuclear binding forces in
a few-nucleon system. This conclusion, of course, does not depend
on the form which one uses for the approximation of $r^A(x)$. However,
the number of parameters used for the approximation
may be critical for the understanding of the modifications pattern
if the experimental errors in $r^A(x)$ are compared with its deviations
for unity.

%
%
 
The magnitudes $m_i$  of distortions of $F_2(x)$ by nuclear 
environment have  been found to  increase monotonously with $A$ and
 to vary similarly~\cite{sm95,sm94} in all the
intervals that used to be regarded as the domains for one 
particular mechanism of the $F_2(x)$ modifications. The $A$ dependence
 of $m_i$ can be approximated in each interval as
\begin{equation}
m_i(A) = M_i (1 - N_s(A)/A), \hspace{1cm} i = 1,2,3,
\label{mwsa}
\end{equation}
where $M_i$ are normalization parameters and $N_s(A)$ is the number of
nucleons on a nuclear surface evaluated with the Woods--Saxon
potential and with  parameters established in the elastic scattering of
electrons off nuclei.
I show below that  Eq.~(\ref{mwsa}) is also valid
for the evaluation of $A$-dependent modifications of $F_2(x)$
beyond $x = 0.7$, within the entire {\em binding} effects
interval 0.3  $< x <$  0.96:
\begin{equation}
m_b(A) = M_b (1 - N_s(A)/A).
\label{bwsa}
\end{equation}

%
%
As has been shown in a number of publications reviewed in
Refs.~\cite{rev2,rev3}, the pattern of the $F_2(x)$ modifications in the
 range 0.3 $< x <$ 1.0
could be qualitatively reproduced with nuclear binding effects and 
Fermi motion corrections. On the other hand, quantitative description
of the $r^{A/{\rm D}}(x)$ within the conventional nuclear structure 
models has been getting worse with improvements of both the data
quality and of the model considerations. The situation has been 
considered as indicating the presense of quark degrees of freedom in heavy
nuclei, which could be used to motivate measurements
 of $F_2^A(x)$ at $x >$~1.
A number of models (c.f. Ref.~\cite{rev3}) were in contradiction with the
frozen pattern of modifications of $F_2(x)$ found from experiment.
Some recent publications have come up with statements that the nucleon
structure is not very much affected by nuclear binding~\cite{vantt}.

At the same time it was clear that the relation between the
EMC effect and nuclear binding effects, recognized by many authors
(c.f. Refs.~\cite{akv}-\cite{bps}), could not be discussed regardless
 of the phenomenon of {\em saturation} 
of nuclear binding forces in the lightest nuclei.
The first attempt to evaluate evolution of $F_2(x)$ in the range
$A \leq$ 4 has been made in Ref.~\cite{bms1} by developing a relativistic
approach for the consideration of nuclear binding effects in $F_2(x)$.
The calculated pattern of the $r^{A=3/D}(x)$ turned out to be similar
to that of $r^{Fe/D}(x)$ determined from experiments, but smaller
in magnitude of deviation of the ratio from unity. This means
that the deviations found in the system $A$~=~3 can be scaled
to $A$~=~56 with $x$-independent parameter $\rho$:
\begin{equation}
1 - F_2^{\rm Fe}(x)/F_2^D(x) ~=~  \rho (1 - F_2^{A=3}(x)/F_2^D(x)) . 
\label{scale}
\end{equation}
The relation has to be considered as a theoretical justification
of the factorization of the $x$ and $A$ dependence 
known from the data analysis of Ref.~\cite{sm95}.
The purpose of my new analysis was to find from experimental data 
the exact pattern of binding effects in the ratio
$F_2^A(x)/F_2^D(x)$ in the entire range of $x$, and to verify how good 
the theoretical calculations for $A \leq$ 4~\cite{bms1} could be
consistent with the available data for $A~\geq$~4. 
I evaluate below the $A$ dependence of  nucleon structure function 
distortions by splitting the range of $x$ in four intervals:\\

\begin{tabular}{llrll}
(1) & {\em nuclear shadowing} \hspace{0.5cm} & $10^{-3}$ &$< x <$ &0.1 ,\\

(2) & {\em antishadowing} region &  0.1 & $< x <$ & 0.3 ,\\

(3) & {\em EMC effect} region &  0.2 & $< x <$ & 0.65,\\

(4) &  {\em nuclear binding} &  0.3 & $< x <$ & 0.96.\\[0.5cm]
\end{tabular}

 Even if there had existed four different mechanisms responsible
for the $x$ and $A$ dependence of $r^{A/{\rm D}}(x)$ in these four
intervals it would have been unlikely that they would have sharp 
boundaries in $x$. Therefore one can allow for an overlap in selection
of the intervals.\\

 {\bf 2.1 Nuclear Shadowing}\\[-0.3cm]

In the range  $x \ll$ 1, which corresponds to nuclear shadowing
 region,  Eq.~(\ref{x1x3}) reduces to 
\begin{equation}
r^{A/{\rm D}}(x)  ~=~  C^{A/{\rm D}} x^{m_{sh}^{A/{\rm D}}},
\label{msh}
\end{equation}

The available data on $r^{A/{\rm D}}(x)$ from EMC~\cite{copper},
NMC~\cite{ama95,arn95} and E665~\cite{xe92,ad95} collaborations are
well approximated with Eq.~(\ref{msh}), demonstrating thus the feasibility
of the factorization of $x$  and $A$ dependence in the shadowing region.
The obtained parameters $m_{sh}^{A/{\rm D}}$ as a function of $A$ are
displayed in Figure~2a.
 Full line is defined by Eq.~(\ref{mwsa}) with $M_{sh}$ = 0.129.

It is also clear that the same pattern holds for the ratio of any
pair of nuclei and therefore the deuteron can be replaced by some
other reference nucleus, for instance by carbon:
\begin{equation}
r^{A/{\rm C}}(x) \equiv F_2^A(x) / F_2^{\rm C}(x)
 = C^{A/{\rm C}} x^{m_{sh}^{A/{\rm C}}} .
\label{mshc}
\end{equation}
%
\begin{center}
\begin{tabular}{ll}
\parbox[t]{8.5cm}{
\begin{minipage}[h]{0.45 \linewidth}
\begin{center}
\mbox{\epsfysize=\hsize\epsffile{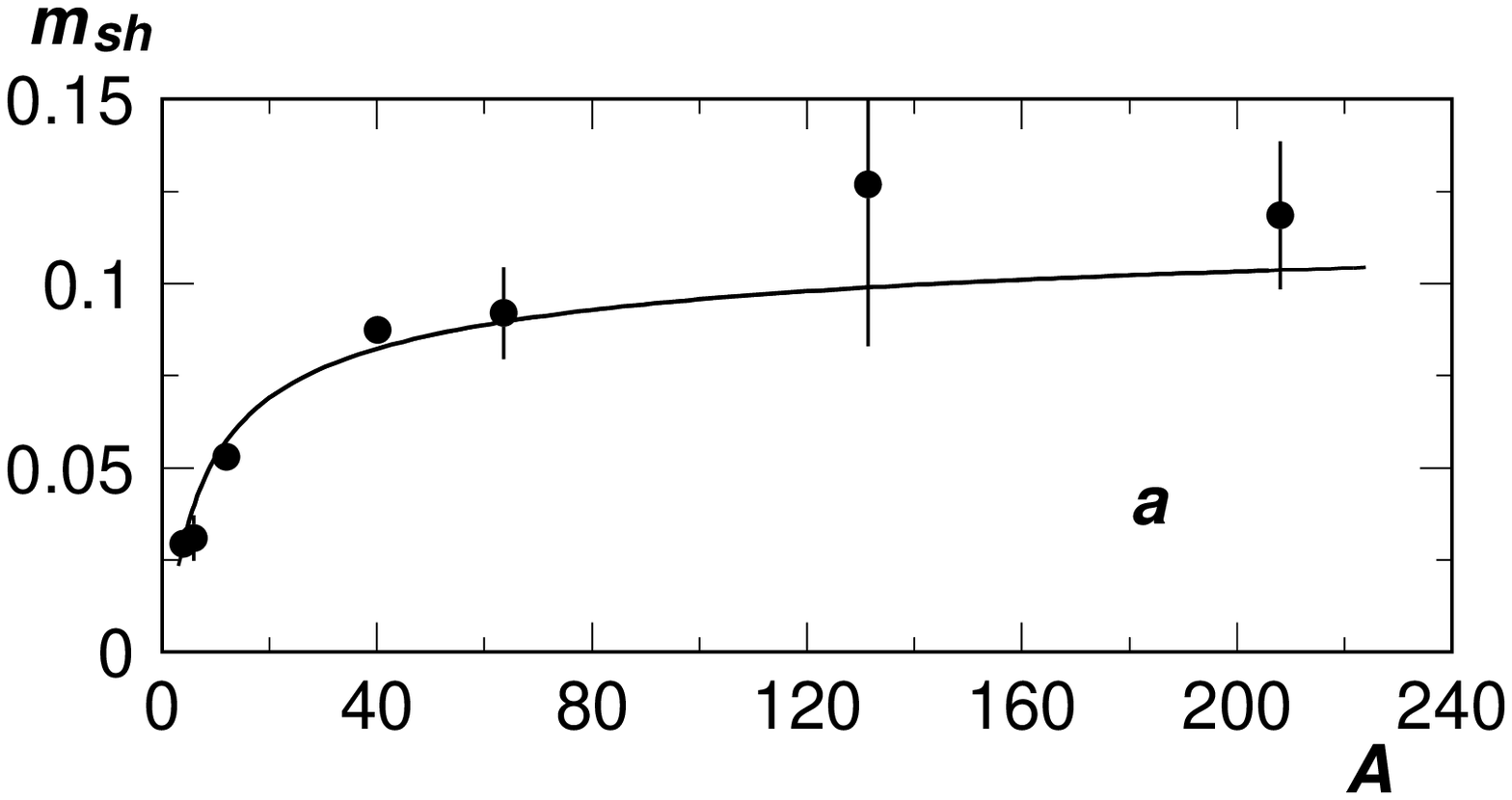}}
\end{center}
\vspace{-2cm}
\begin{center}
\mbox{\epsfysize=\hsize\epsffile{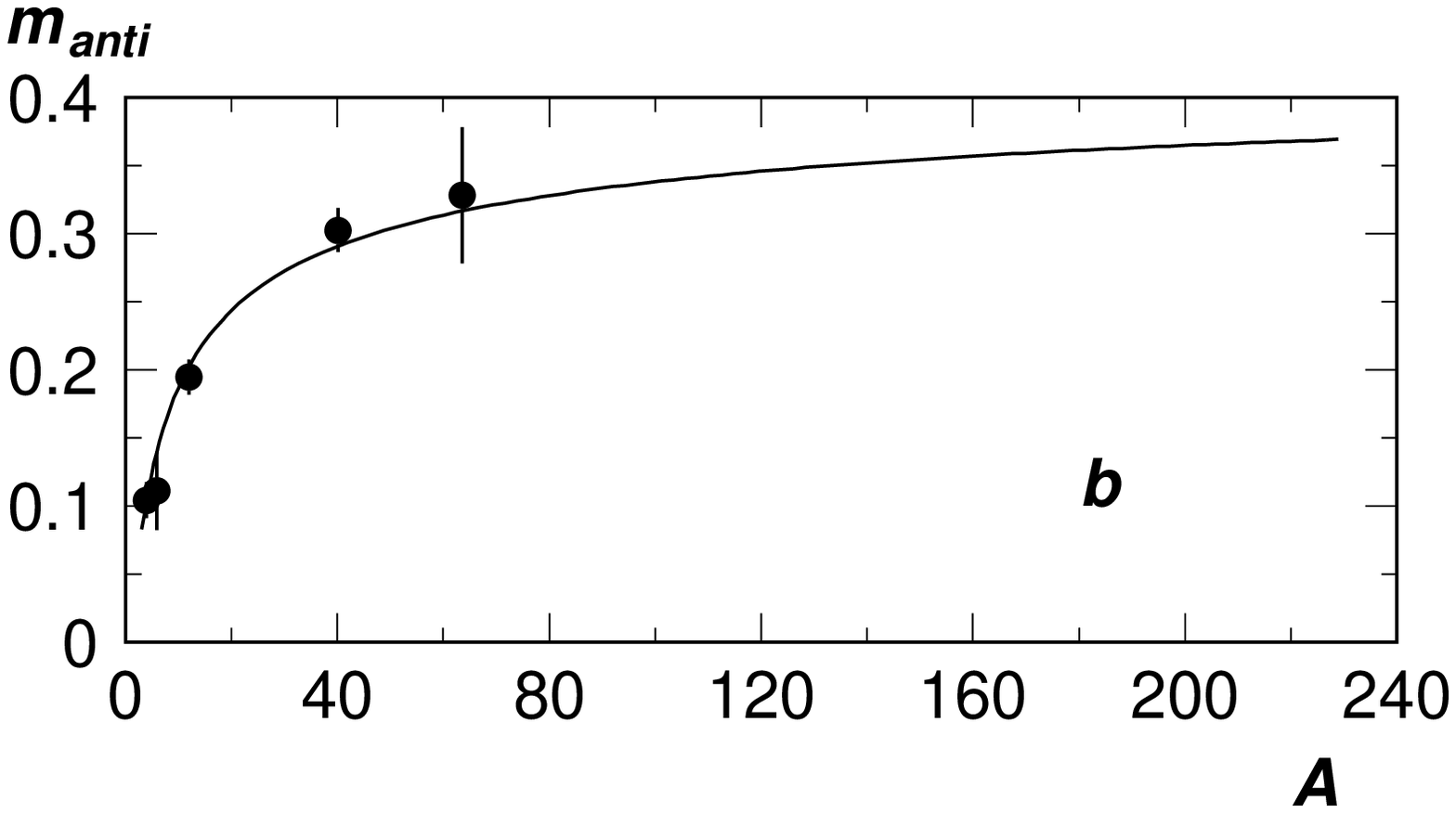}}
\end{center}
\vspace{-2cm}
\begin{center}
\mbox{\epsfysize=\hsize\epsffile{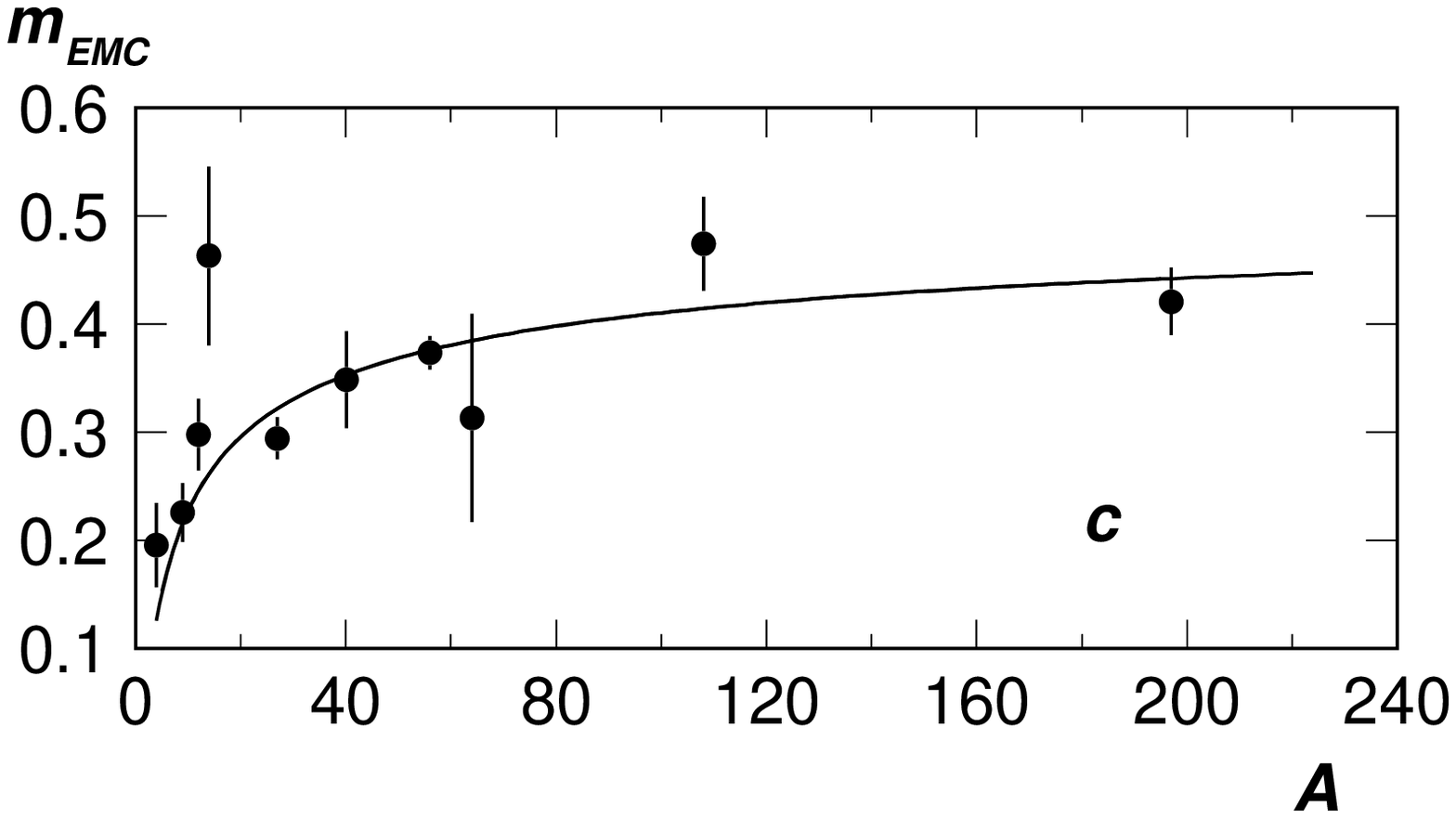}}
\end{center}
\vspace{-2cm}
\begin{center}
\mbox{\epsfysize=\hsize\epsffile{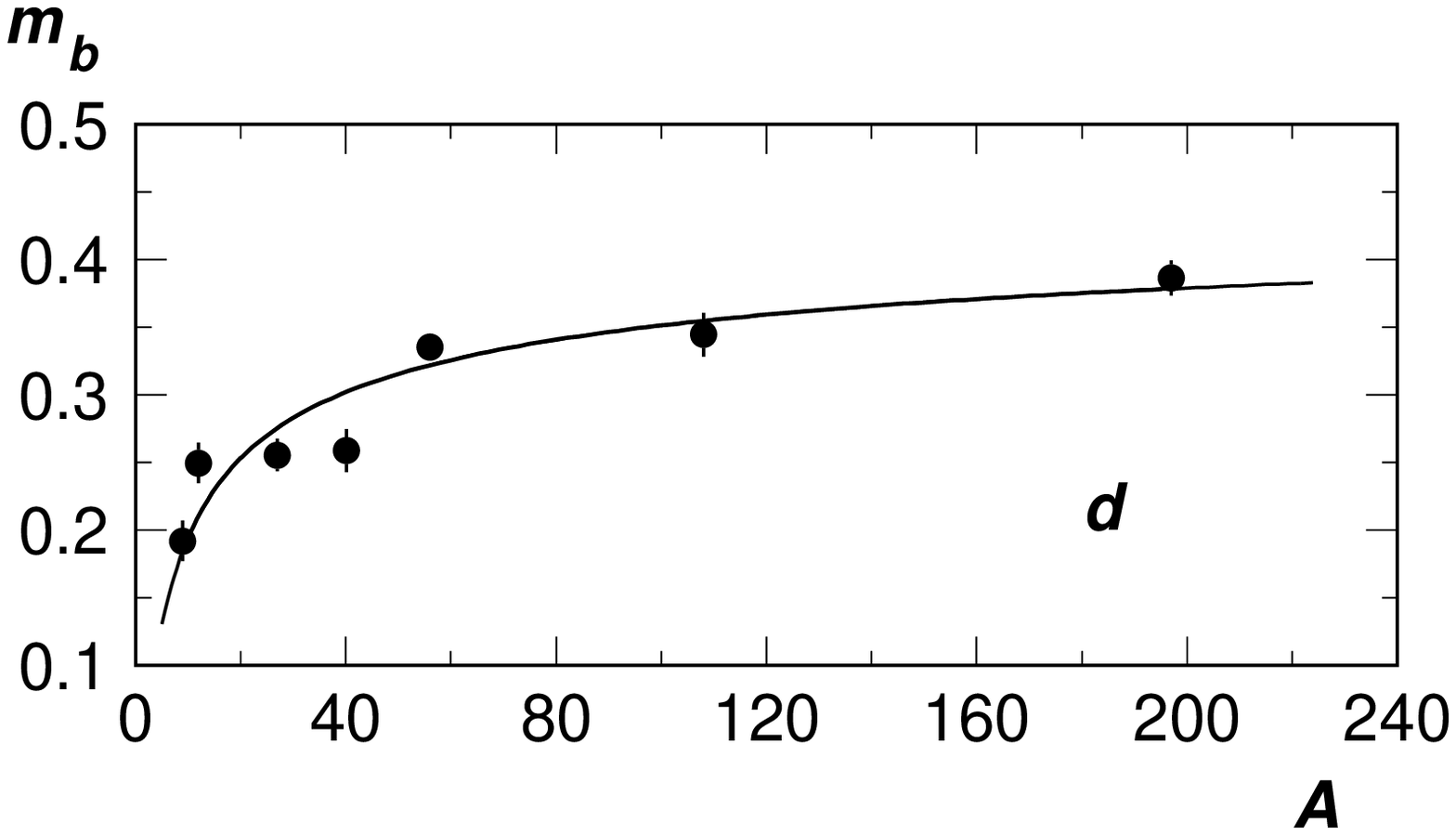}}
\end{center}
\end{minipage}}&
\parbox[t]{6.5cm}{
Figure 2.  The parameters {\boldmath $m$}, which
 define the magnitude of distorsions
of $F_2(x)$, determined in the  regions of nuclear shadowing (a),
antishadowing (b), EMC effect (c) and in the high $x$-range (d).
Full lines in (a) -- (c) are obtained with Eq.~(2) and in (d) with
Eq.~(3). The number of nucleons $N_{\rm s}(A)$ at the nuclear
surface is given by the Woods--Saxon potential:\\[-0.3cm]

$N_{\rm S}(A) = 4 \pi \rho_0 \int \limits_{r_0(A)}^{\infty}
{ {dr r^2} \over {1 + e^{[r-r_0(A)]} / a} }$.
}
\end{tabular}
\end{center}
\setcounter{figure}{2}

I find that the recent NMC results on the
structure functions ratios  measured on
Be, Al, Ca, Fe, Sn and Pb targets  with respect to carbon~\cite{arn96}
are well approximated with Eq.~(\ref{mshc}). 
From the comparison of Eq.~(\ref{msh}) and Eq.~(\ref{mshc}) I obtain
the relation between the distortion magnitudes $m_{sh}$ determined
from $A/\rm D$ and $A/\rm C$ data:
\begin{equation}
m_{sh}^{A/{\rm D}} = m_{sh}^{A/{\rm C}} + m_{sh}^{\rm C/D} ,
\label{alphad}
\end{equation}
I apply Eq.~(\ref{alphad}) to the
distortion parameters $m_{sh}^{A/{\rm C}}$ evaluated from the
data of Ref.~\cite{arn96} and plot the results in Figure~3  together with
the results of direct determination of $m_{sh}^{A/{\rm D}}$. Larger
errors from the $A/{\rm C}$ experiment are explained by a considerably
larger nuclear shadowing effect in carbon nucleus, which results
in smaller differences between cross-sections measured on nuclear
targets and on a carbon target. Within the experimental errors 
both experiments are consistent.\\
%
%
%
\begin{figure}[h]
\begin{center}
\mbox{\epsfxsize=0.5\hsize\epsffile{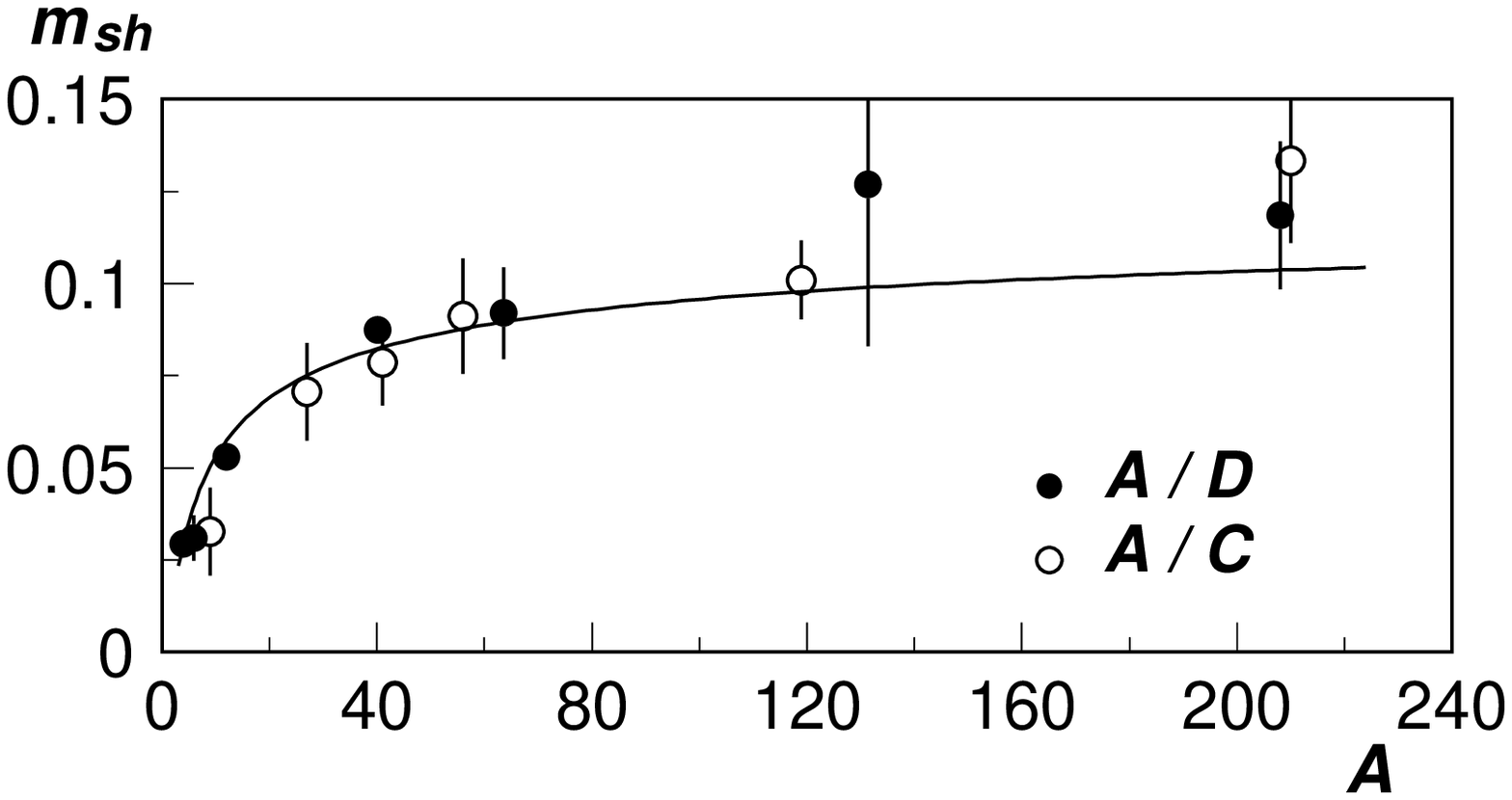}}
\end{center}
\vspace{-0.5cm}
\caption{ The parameter $m_{sh}$ evaluated from the data on
$F_2^A(x)/F_2^D(x)$ (full circles) and from $F_2^A(x)/F_2^C(x)$
(open circles). Full line  is defined by Eq.~(2)
}
\end{figure}

 {\bf 2.2 Antishadowing}\\[-0.3cm]

As follows from the chosen form of the approximation function Eq.~(\ref{x1x3}),
the deviations of the parameter $m_{anti}^{A/{\rm D}}$ from zero would
 detect nuclear medium effects
 in the {\em antishadowing} region. Antishadowing has not been
studied so far quantitatively because of evident problems of the
measurements of the effect which is comparable with
experimental errors. This is why one can not rely on the data
which does not cover a considerably wider range  than 0.1 $< x <$ 0.3.  
This concerns the data of BCDMS~\cite{bcdms}, E665~\cite{ad95} 
and also the SLAC data~\cite{gomez} for some targets (Be, Al, Fe, Ag, Au).
On the other hand, the SLAC and NMC~\cite{ama95} data on $^4$He, C and Ca
combined together cover nearly full $x$ range and are very well suited
for the studies of the small antishadowing effect. Equally good
proved to be the data of NMC~\cite{arn95} for Li and of EMC~\cite{copper}
for Cu targets.
The obtained parameters $m_{anti}^{A/{\rm D}}$ as a function of $A$
are displayed in Figure~2b. 
Full line is defined by Eq.~(\ref{mwsa}) with $M_{anti}$ = 0.456.\\

 {\bf 2.3 EMC Effect Region}\\[-0.3cm]

As will be shown below, the physics of modifications of $F_2(x)$ in
this range of $x$ is understood as nuclear binding effects. 
Still for the moment I consider 
the data in the region 0.25 $<x<$ 0.65 separately from the
high $x$ range because the largest number of data has been collected
in this very region (10 nuclear targets) and they can all be reasonably
well approximated with a linear equation,

\begin{equation}
r^{A/{\rm D}}(x)  ~=~  a - {m_{\mbox{\tiny EMC}}}x . 
\label{memc}
\end{equation}

The obtained parameters $m_{\mbox{\tiny EMC}}^{A/{\rm D}}$ as a function
of $A$ are displayed in Figure~2c.  Full line is defined
by Eq.~(\ref{mwsa}) with $M_{\mbox{\tiny EMC}}$ = 0.553.\\

 {\bf 2.4 Nuclear Binding}\\[-0.3cm]

 As has been obtained in Ref.~\cite{bms1}, modifications of the
$x$ dependence of $F_2(x)$ result from the nuclear binding and
are the strongest in the four-nucleon system, $^4$He.
Modifications predicted for the three-nucleon system were found to be
identical in form and different in the amplitude from those
experimentally observed in heavy nuclei. To verify the latter
statement I introduce, below, two equations for approximation
of the data in the range $x >$ 0.3,
\begin{eqnarray}
\label{mb3}r^{A/{\rm D}}(x)&&=~~  1 - m_{b}(A) a_{osc}^{A=3}(x),
 \hspace{1cm} A \neq 4 ,\\
\label{mb4}r^{A=4/{\rm D}}(x)&&=~~  1 - {m_{b}(A=4)} a_{osc}^{A=4}(x),
\end{eqnarray}
where $m_{b}(A)$ is a free parameter, ${m_{b}(A=4)}$ = 0.24,
and $a_{osc}^{A=3(4)}(x)$ is defined as a relative difference between
the structure functions
of the 3(4)-nucleon system  $F_2^{A=3(4)}(x)$ and that of the deuteron:
\begin{equation}
a_{osc}^{A=3(4)}(x) \equiv 1 ~-~ F_2^{A=3(4)}(x) / F_2^{\rm D}(x) . 
\label{a34}
\end{equation}
\setcounter{figure}{3}
%
\begin{figure}[t]
\begin{minipage}[t]{0.34\linewidth}
\begin{center}
\mbox{\epsfysize=\hsize\epsffile{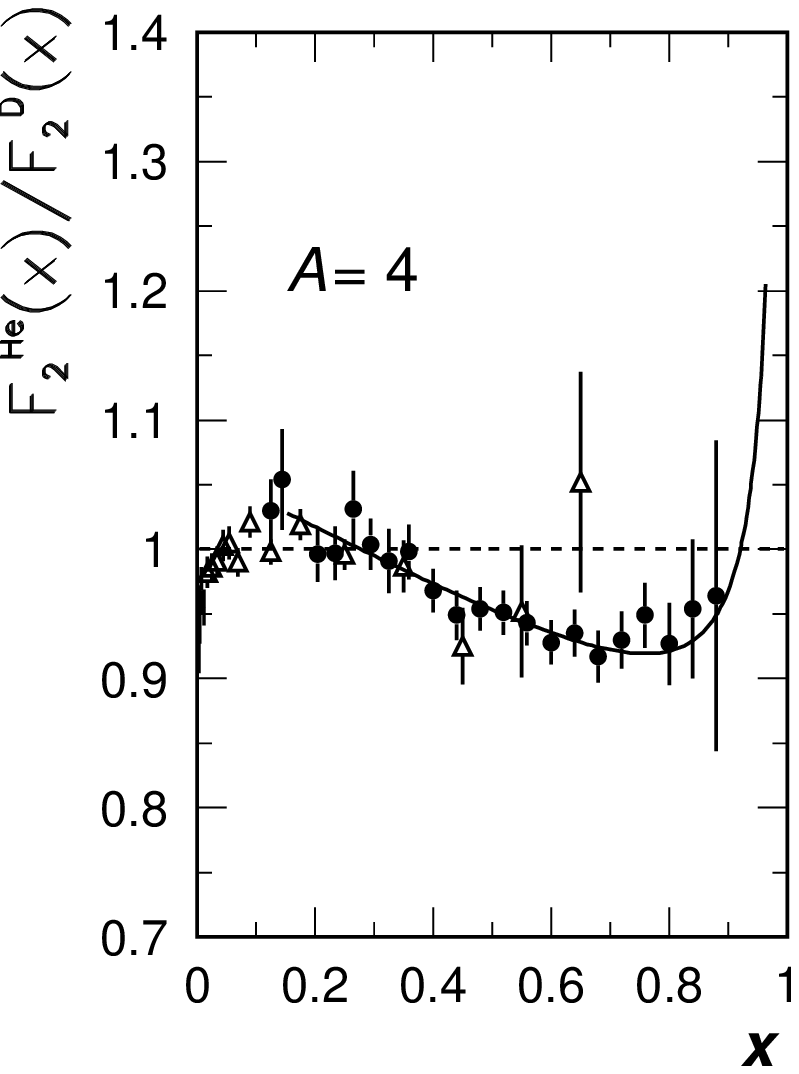}
\hspace{-0.4cm}\epsfysize=\hsize\epsffile{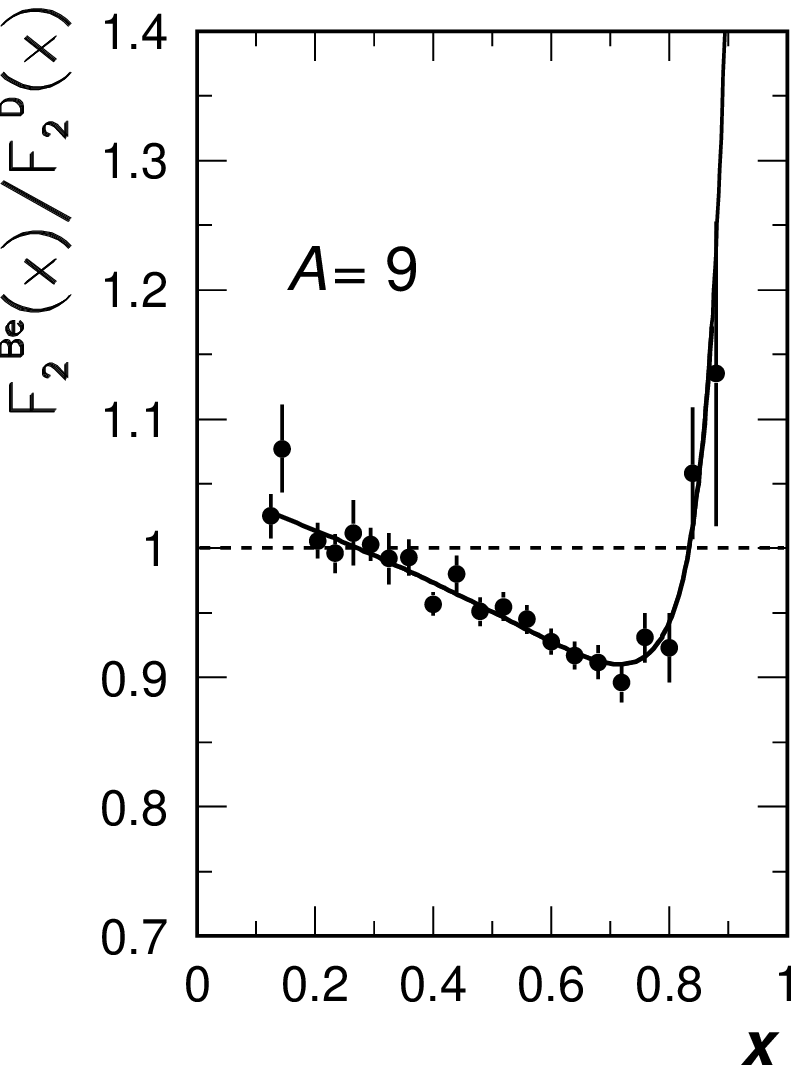}
\hspace{-0.4cm}\epsfysize=\hsize\epsffile{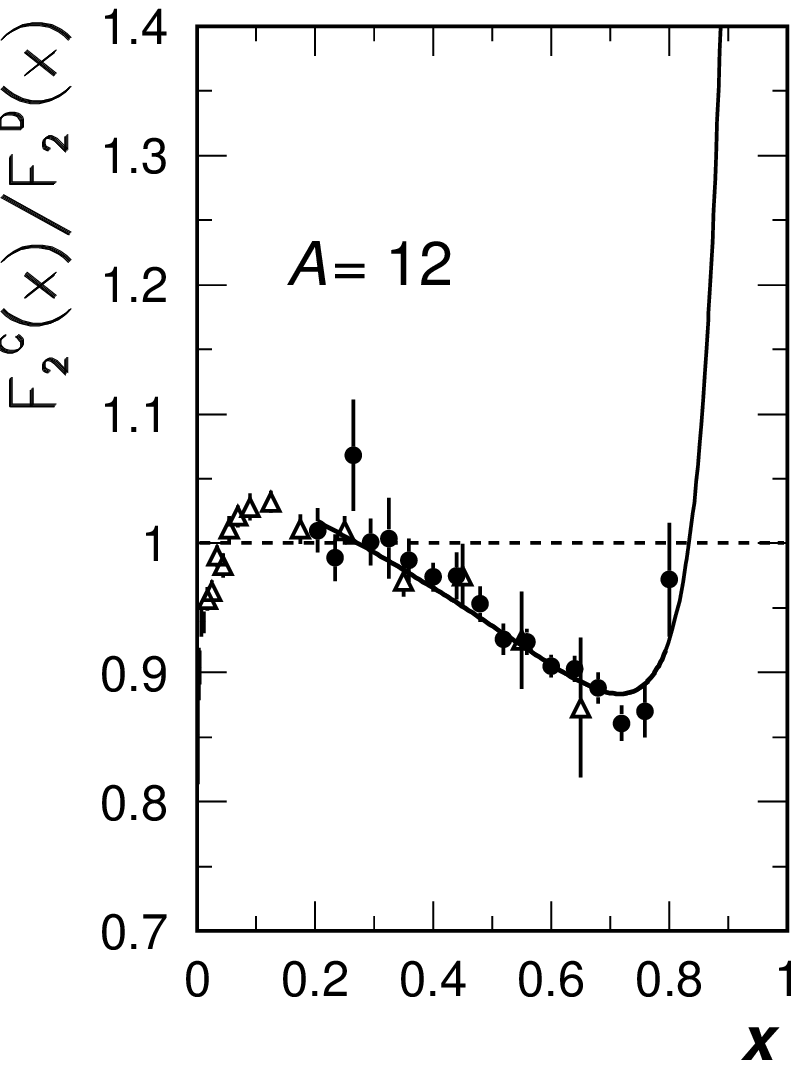}
\hspace{-0.4cm}\epsfysize=\hsize\epsffile{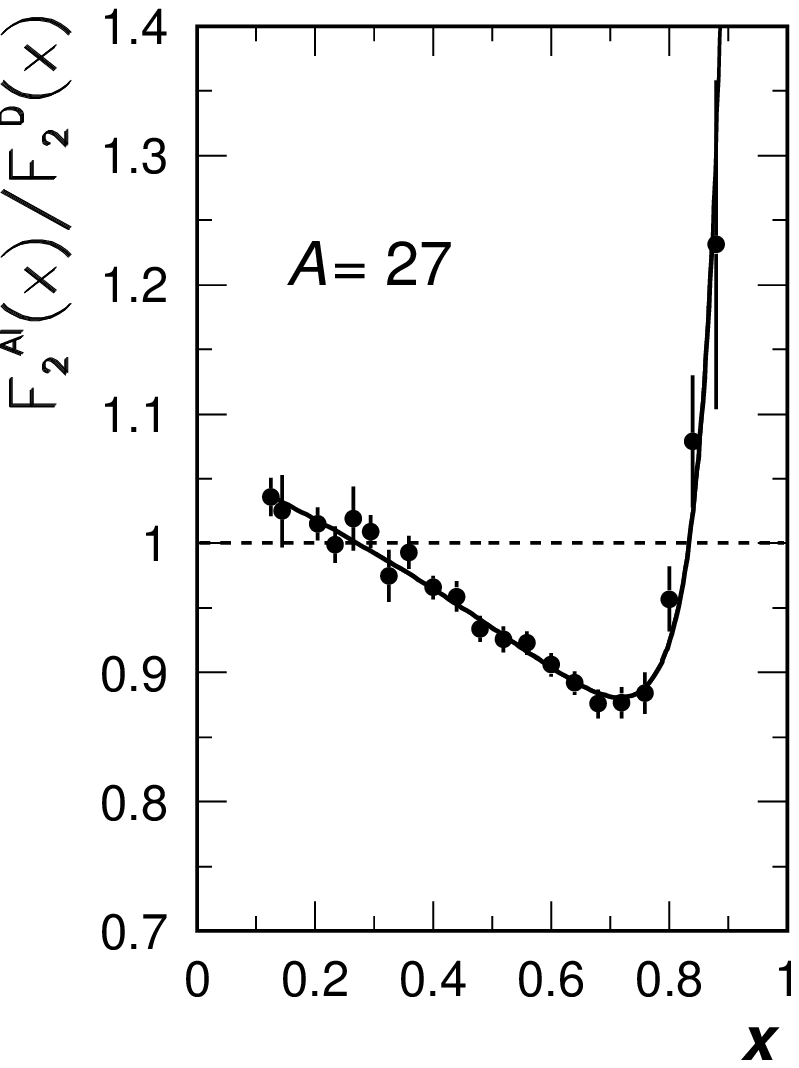}}
\end{center}
\end{minipage}
\vspace{0.2cm}

\begin{minipage}[t]{0.34\linewidth}
\begin{center}
\mbox{\epsfysize=\hsize\epsffile{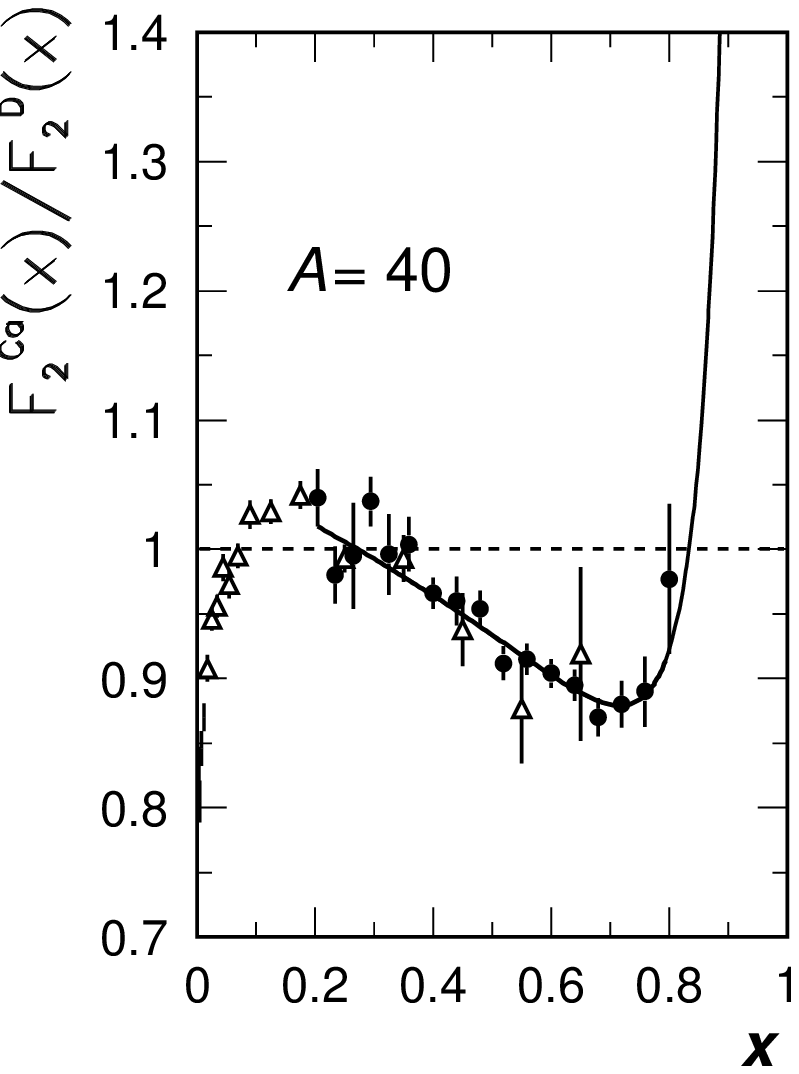}
\hspace{-0.4cm}\epsfysize=\hsize\epsffile{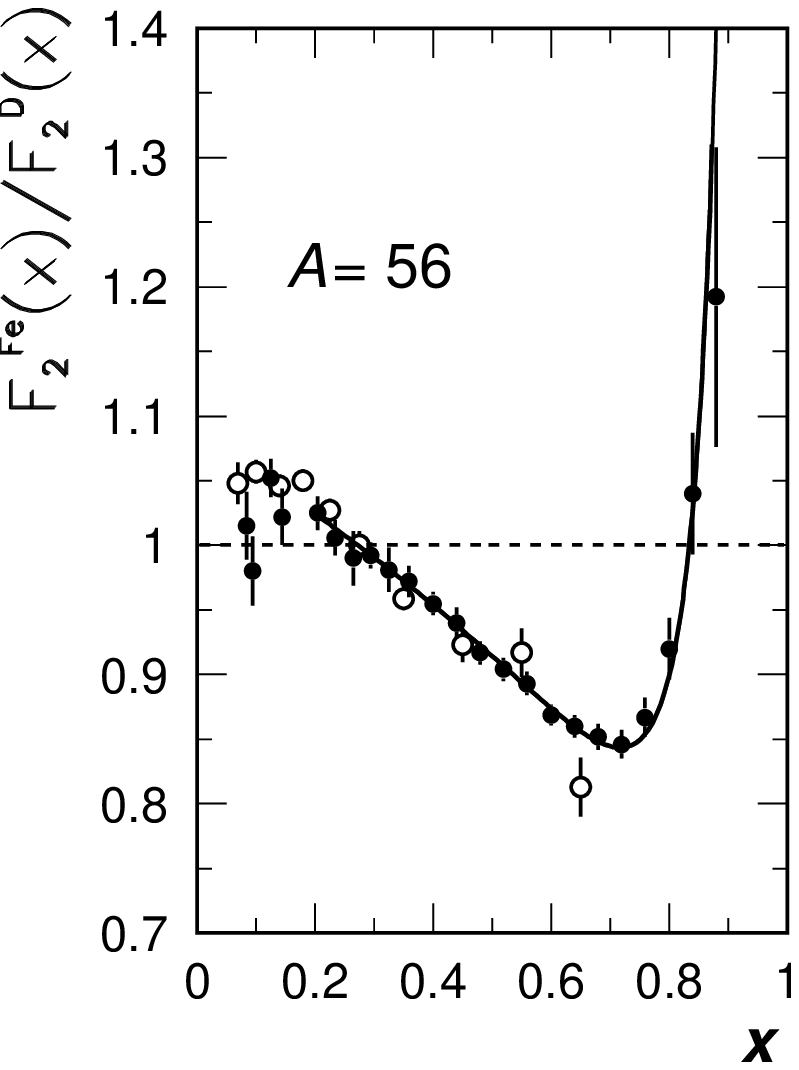}
\hspace{-0.4cm}\epsfysize=\hsize\epsffile{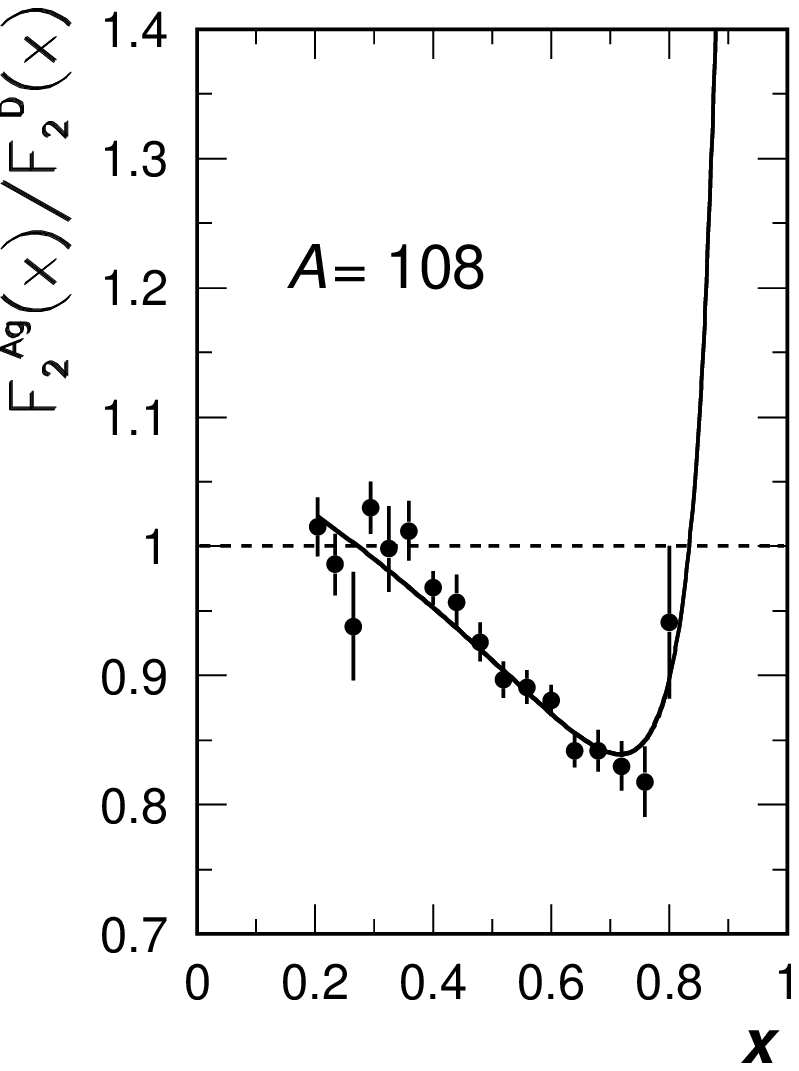}
\hspace{-0.4cm}\epsfysize=\hsize\epsffile{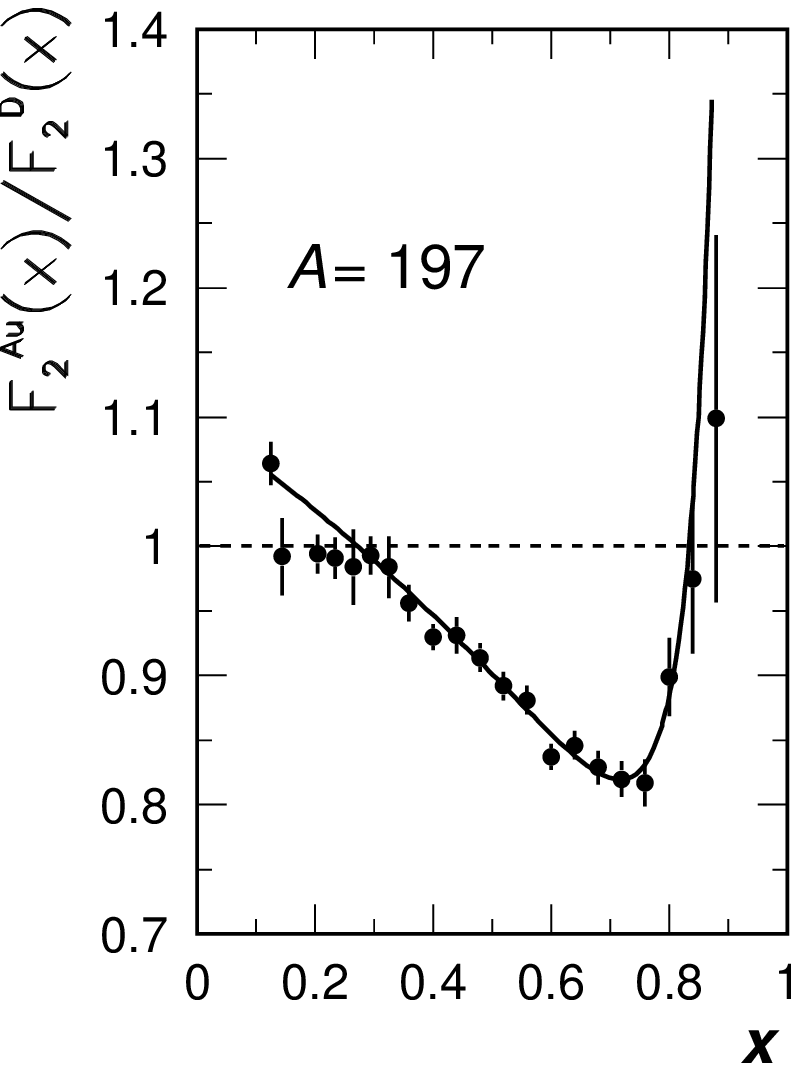}}
\end{center}
\end{minipage}
\vspace{0.2cm}
\caption{ 
 Comparison  of $F_2^A/F_2^{\rm D}$,
measured by SLAC, BCDMS and NMC, in the range $x>$ 0.2,
with theoretical calculations~[5]
for  $^4$He ( $A$ = 4) and $^3$He  ( $A \geq$ 9).
Only one normalization parameter $m_b(A)$ is used to adjust
theoretical results for $^3$He with the data.}
\end{figure}

The evolution of the isoscalar nucleon structure function
$ F_2^{\rm N}(x)$ from $A$~=~1 to $A$~=~4,
according to Ref.~\cite{bms1}, is defined largely by a series of
terms containing derivatives of 
$ F_2^{\rm N}(x)$, $ F_2^{\rm D}(x)$ and $ F_2^{A=3}(x)$.
In the simplest case of the input $F_2^{\rm N}(x) \sim (1-x)^3$ 
the modifications  are represented  as
a power series of $1/(1-x)$ terms. Applying the well established
boundary condition $a_{osc}(x_2)$ = 0, I obtain a simple analytical
equation which describes modifications of parton distributions
caused by binding forces in the lightest nuclei:

\begin{equation}
a_{osc}^{A=3(4)}(x) = \biggl(1-\lambda^{A=3(4)} x \biggr)
 \Biggl\{ \Biggl( \frac{1}{u} - \frac{1}{c} \Biggr)
~-~ \mu^{A=3(4)} \Biggl( \frac{1}{u^2} - \frac{1}{c^2} \Biggr) \Biggr\}, 
\label{osc2}
\end{equation}
where $u = 1 - x$, $c = 1 - x_2$, $\lambda^{A=3(4)}$ = 0.5 (1.0).
 Parameter $\mu^A$ is defined by
the requirement $a_{osc}(x_3)$ = 0 and its numerical value obtained
in Ref.~\cite{bms1} corresponds to $\mu^{A=3(4)}$ =
${m_{\pi}}/{\rm {M}} ({m_{\pi}}/{\rm {2M}})$,
where $m_{\pi}$ and ${\rm M}$ are the pion and nucleon masses.

It is important to note that Eq.~(\ref{osc2}) does not contain any free
parameter except $c$, which is constrained with experimental results
for $x_2$ and, as shown below, can also be expressed through the 
value of $x_3$. When evaluated with Eq.~(\ref{osc2}), 
$a_{osc}^A(x)$ virtually coincides with numerical values
 of Ref.~\cite{bms1} for $x>$ 0.3.\\

As obtained in Ref.~\cite{bms1}, the coordinate $x_3$ for $A$ = 4 is
 twice as close to the kinematic boundary as that for $A$ = 3:
($1 - x_3^{A=3}$)/($1 - x_3^{A=4}$) $\approx 2$,
which is reflected in the relation between
the parameters $\mu^{A=3}$ and $\mu^{A=4}$ of Eq.~(\ref{osc2}).
This makes the pattern of distortions for $^4$He different from 
the rest of nuclei.  It is compared with data in Figure~4.
Experimental results for $A >$ 4
in Figure~4 are approximated with Eq.~(\ref{mb3}) with one free
parameter $m_{b}(A)$. The results of approximation are displayed
in Figure~4 as a function of $x$ and in Figure~2d as a function of~$A$.
Full line in Figure~2d is defined by Eq.~(\ref{bwsa}) with $M_b$ = 0.473.
From good agreement between the theory and data, which is evident from
Figure~4, I find
that $x$ dependence of deviations from $r^{A/{\rm D}}(x)$ =~1
remains unmodified in the entire range of atomic weights~$A$
and is well described by {\em scaling} the amplitude $a_{osc}$ of
deviations evaluated for  $A$ = 3. 
This means that  modifications of $F_2(x)$ in heavy nuclei
{\em saturate} even faster than in the lightest nuclei.

   As demonstrated with Figures~1 -- 4, the $x$ and $A$ dependence
of the modifications can be factorized in the entire range of $x$. The 
phenomenon is nicely reproduced with Eqs.~(\ref{x1x3}) and~(\ref{mwsa})
in the range $x < 0.7$ and with Eqs.~(\ref{mb3}) and~(\ref{bwsa})
in the range $x > 0.3$. This gives one a simple tool to plot
the two-dimensional pattern of modifications of the nucleon 
structure function in nuclear environment which is shown in
Figure~5. It should be underlined that the $A$ dependent evolution
represented by the plot is largerly the result of the variation of
the nuclear surface-to-volume ratio, while the evolution of
the  partonic distribution remains with the lightest
nuclei, $A \leq$ 4.
%
%
\begin{center}
\begin{figure}[t]
\begin{center}
\begin{minipage}[t]{0.5 \linewidth}
\hspace*{-4cm}\mbox{\epsfysize=\hsize\epsffile{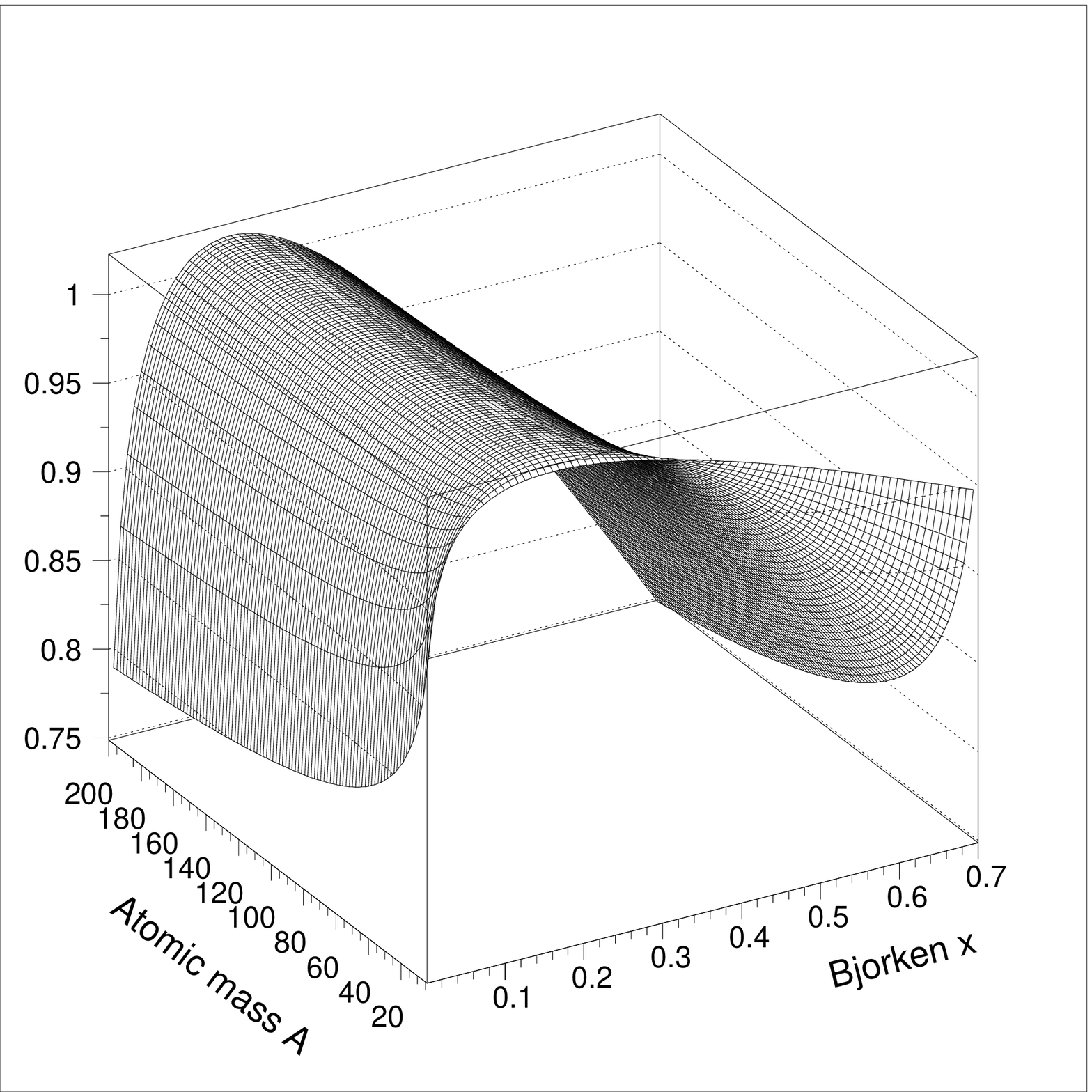}
\epsfysize=\hsize\epsffile{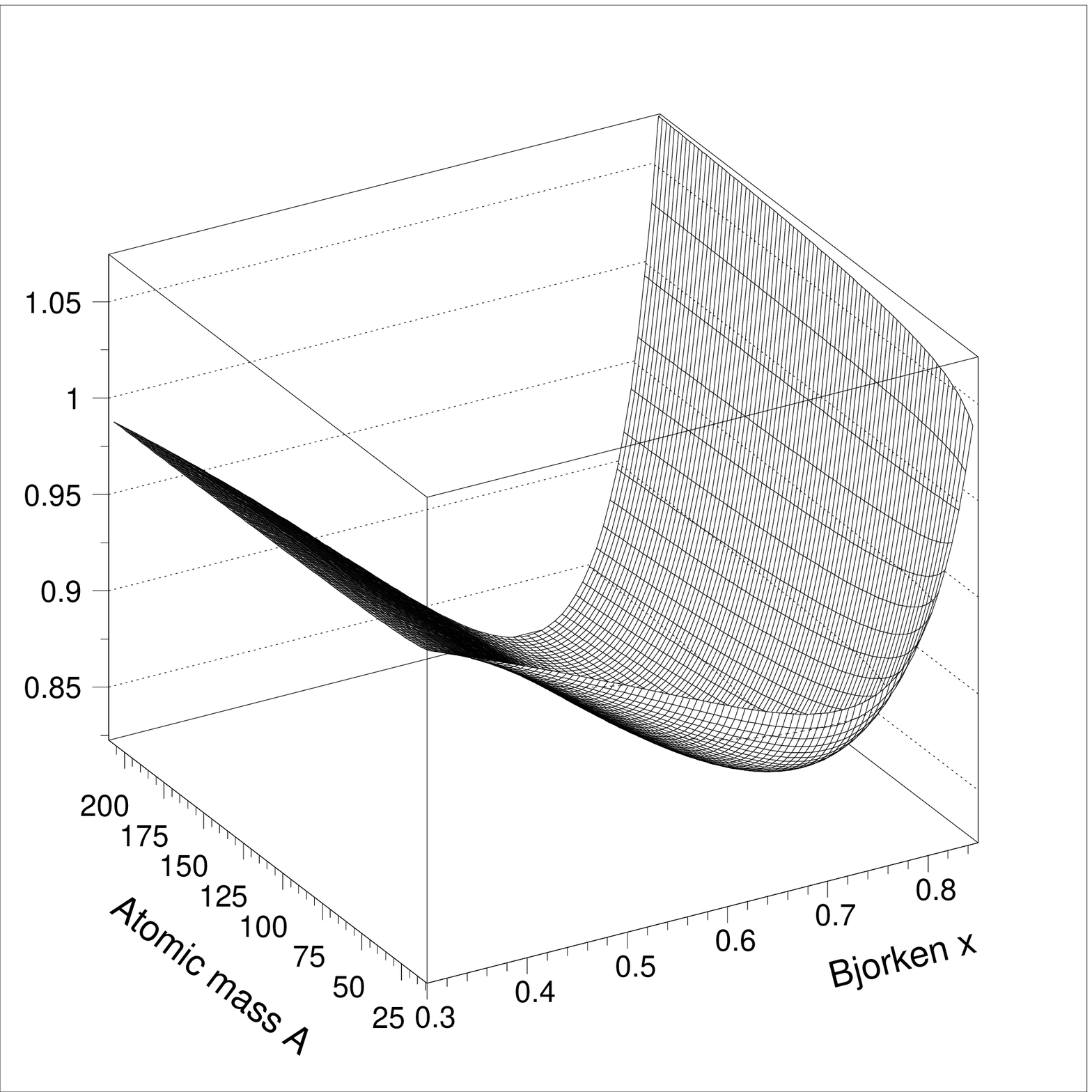}}
\end{minipage}
\end{center}
\caption{ 
Approximation of the pattern of the EMC effect as a function
of $x$ and $A$ in the range 1) $x <$ 0.7, $A \geq 4$ (left frame)
and 2) $x >$ 0.3, $A \geq 9$ (right frame).}
\end{figure}
\end{center}

The pattern shown in Figure~5 is obtained for the measured range of
$x$ and $A$ only. Its extrapolation to larger values of $x$ and
$A$ can be justified by the consistency between 
the experimental data analysis and calculations of Ref.~\cite{bms1}.\\

{\bf 3~ Role of the Partonic Structure in the Pattern of Binding}\\[-0.3cm]

An important feature of the factorization of the $x$ and $A$
dependence of the modifications in the range $A >$ 4,
is the $A$ independence of the coordinates of
the three cross-over points $x_i$. There are reasons
to believe that they are constrained by the inner structure of
the nucleon and therefore are strongly correlated. Nevertheless there
exists rich literature which discusses the role of different mechanisms
responsible for the nuclear effects in the low and high $x$ range
and insists that the coordinates must be considered as unrelated with
each other. This motivated the tests of the $A$ independence of
$x_1$~\cite{sm94} and $x_2$~\cite{sm95}. Below I refresh 
experimental status of the coordinates $x_{i=1,2}$ and present new
results of determination of $x_3$. The latter can now be compared with 
theoretical calculations of Ref.~\cite{bms1}. 

There is a definite advantage to relate $x_i$ with the pattern of 
$r^{A/{\rm D}}(x)$ because of the two reasons:

1) the coordinates $x_i$ are much less dependent on the form of
 approximation functions, which makes them more sensitive to possible
 $A$ dependence than the functions themselves, and 

2) the coordinates $x_i$ can be easily obtained as fully independent
 from each other in the space of Bjorken variable $x$, which is important
 for the understanding of which physics is responsible for the pattern.\\

 {\bf 3.1 First cross-over}\\[-0.3cm]

I find $x_1$ as an intersection point of a straight line
$r^{A/{\rm D}}(x)$  = 1 with  $r^{A/{\rm D}}(x)$ given by Eq.~(\ref{msh}).
The parameters $C$ and $m_{sh}$ have been found by fitting the data
in the range 0.001 $< x <$ 0.08 on He, Li, C and Ca obtained
 by NMC~\cite{ama95,arn95}, on Cu by EMC~\cite{copper} and on 
Xe~\cite{xe92} and Pb~\cite{ad95} by E665. The value $x_1$ as a
function of $A$ is plotted in Figure~6a. Within experimental errors the
 results are consistent with $x_1$ = const ($\chi^2$/d.o.f. = 6.1/7)
and correspond to $\overline {x_1}$ = 0.0615 $\pm$ 0.0024.\\

 {\bf 3.2 Second cross-over}\\[-0.3cm]

I used the same data sample to obtain the coordinate of the second
cross-over point $x_2$ as for determination of $m_{\mbox{\tiny EMC}}$.
It is found as an intersection point of a straight line 
$r^{A/{\rm D}}(x)$  = 1 with $r^{A/{\rm D}}(x)$ given by Eq.~(\ref{memc}):
\begin{equation}
x_2(A) = (a(A) - 1) / m_{\mbox{\tiny EMC}} . 
\label{xemc}
\end{equation}
 The results are plotted in Figure~6b. As in the case for $x_1$ I find
that $x_2$ =  const ($\chi^2$/d.o.f. = 7.4/9).
The mean value is shown with the dashed line
and corresponds to $\overline {x_2}$ = 0.278 $\pm$ 0.008.\\

 {\bf 3.3 Third cross-over}\\[-0.3cm]

The experimental results for the third cross-over point $x_3$ play a
decisive role in the understanding of the pattern of binding effects
in $F_2(x)$. 
 Since there is little data available above $x_3$ one has 
to find some reasonable approximation function in the range 
$x >$ 0.3 to avoid correlations
between data collected on different nuclear targets and between
coordinates of $x_2$ and $x_3$. I have chosen
the function with four free parameters $a_{i=1 - 4}$ as follows:
\begin{equation}
r^{A/{\rm D}}(x) ~=~ a_1 (a_2 - x)
{exp(- a_3 x^2) \over (1 - a_4 x)^{2 - a_1}~}.
\label{x3}
\end{equation}
The results of determination of $x_3$ are plotted in Figure~6c as
closed circles. Again I find that $x_3$ is independent of $A$
within experimental errors ($\chi^2$/d.o.f. = 1.9/6).
The mean value is shown with
the dashed line and corresponds to $\overline {x_3}$ = 0.84 $\pm$ 0.01.
In the same plot I show results of theoretical calculations~\cite{bms1}
for the three- and four-nucleon systems.\\[-0.2cm]

Two important conclusions follow from the obtained results:

1) One finds that the three determined coordinates are fairly well
correlated, namely:
\begin{eqnarray}
 \overline {x_1}  + \overline {x_2} & \approx & ~~1/3,\\
 \overline {x_2} & \approx & ~\overline {x_3} / 3,\\
 \overline {x_3} & \approx & ~~5/6.
\label{x123}
\end{eqnarray} 
The relations (15) -- (17) might play a fundamental role in
understanding both the free nucleon partonic structure and
the mechanism of its modification in nuclear environment.
In particular, Eq.~(16) establishes the relationship between the
theoretically defined  $x_3$ and the still poorly understood $x_2$.
  The precise value of $x_2$ has not been
critical for the theory of Ref.~\cite{bms1} which considered the range 
$x >$ 0.3. On the other hand it has been helpful in bringing the
theory to better agreement with data when Eq.~(\ref{osc2}) was used.
The employment of Eq.~(16) allows one to get rid of free parameters 
in Eq.~(\ref{osc2}). 

%
\begin{tabular}{ll}
\parbox[h]{7.4cm}{
\begin{minipage}[t]{0.7 \linewidth}
\begin{center}
\mbox{\epsfysize=\hsize\epsffile{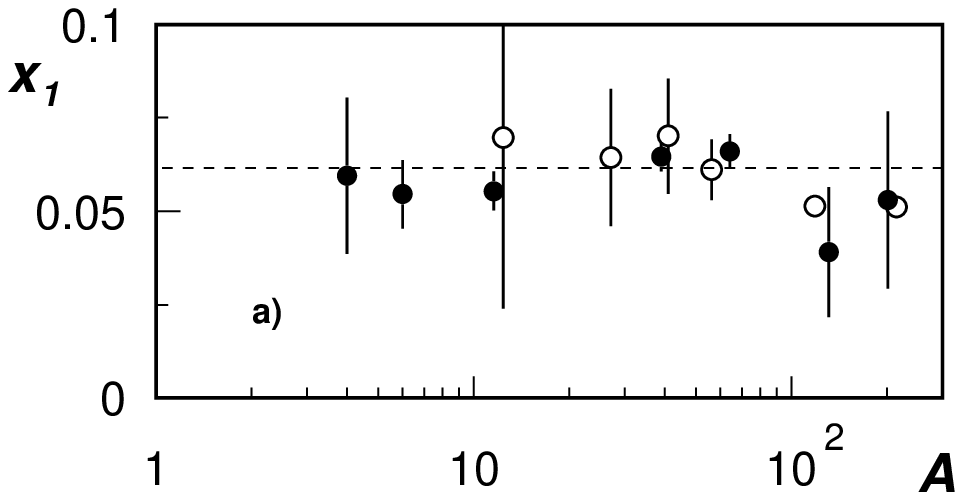}}
\end{center}
\vspace{-3cm}
\begin{center}
\mbox{\epsfysize=\hsize\epsffile{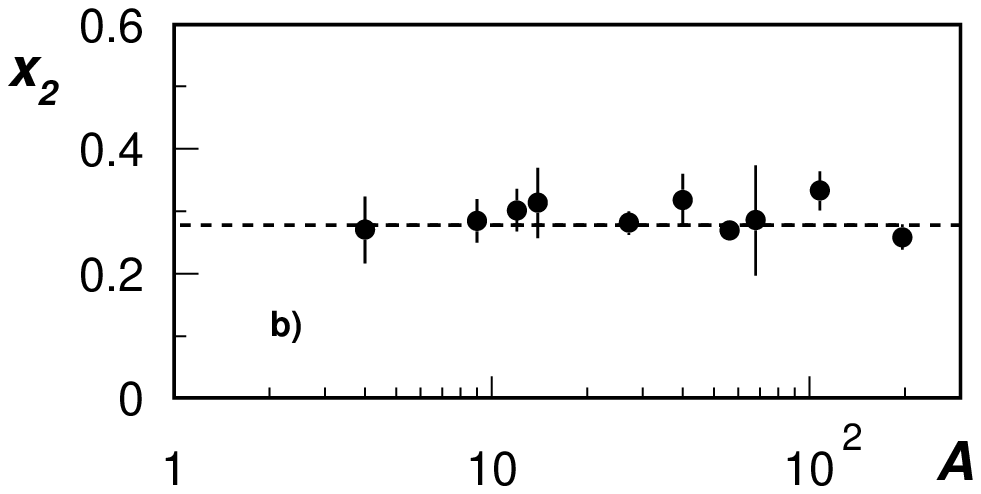}}
\end{center}
\vspace{-3cm}
\begin{center}
\mbox{\epsfysize=\hsize\epsffile{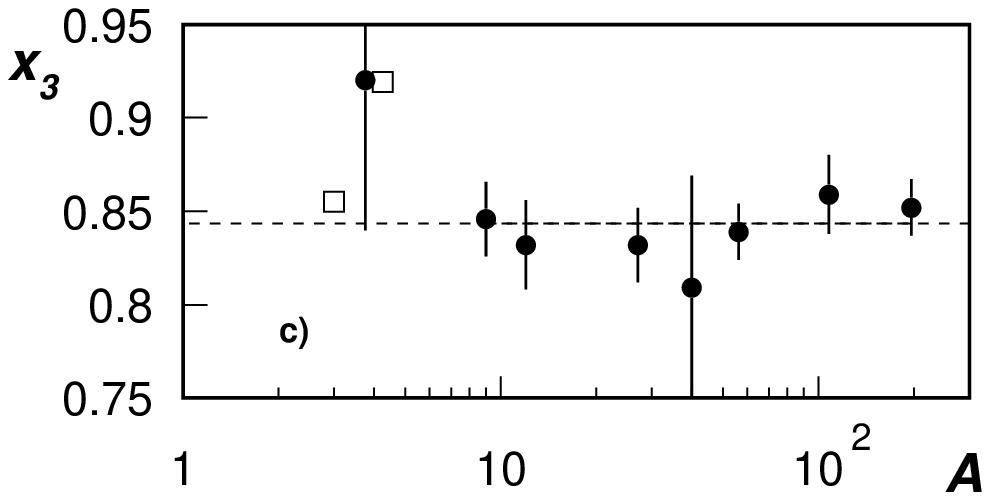}}
\end{center}
\end{minipage}}&
\parbox[h]{7.5cm}{Figure 6. The coordinates of the cross-over
points $x_1$ --- a), 
$x_2$ --- b) and $x_3$ --- c) as a function
of atomic mass $A$. The data on $F_2^A/F_2^{\rm D}$ have been
used to obtain results plotted with full circles. The results
for $x_1$ shown in a) with open circles have been obtained
from the data on $F_2^A/F_2^{\rm C}$.
The average values are shown with the dashed
lines: $\bar x_1$ = 0.0615, $\bar x_2$ =  0.278, $\bar x_3$ = 0.84.
Theoretical values for $x_3$ evaluated for $A$~= 3 and 4
are shown with empty squares.
}
\end{tabular}

2) The $A$ independence of the three cross-over points  serves
as the evidence, that the pattern of distortions can not
be related with properties of the nuclear medium. On the
contrary, it is a message about the nucleon structure which reveals
itself in the presence of binding interactions in a few nucleon 
system.\\

{\bf 4~ Discussion}\\[-0.2cm]

%
%
   The analysis of the world data on the structure function
ratios performed in this paper has demonstrated that the relativistic
theory of nuclear binding~\cite{bms1} is in very good agreement with
experiment. I observe also that the agreement is considerably better
than that obtained by recent explanations of the EMC effect in the QCD
inspired model~\cite{yang} or in the phenomenological double
 $Q^2$-rescaling model~\cite{zhe}.

The new precise picture of the $x$ and $A$ dependence, which stems
from this analysis, serves to explain the observed evolution of
$F_2(x)$ in nuclei with the variation of nuclear density and geometry
of a nucleus in agreement with Woods--Saxon potential. The evolution
does not modify partonic distributions if $A >$ 4 which is
particularly important for the understanding the role of nuclear
environment. It follows therefore  that two options only 
are left for the explanation of the EMC effect origin: 
it is either $F_2^{A=4}(x)$ or $F_2^{A=3}(x)$ which is 
different from $F_2^{\rm D}(x)$ due to nuclear binding effects.   
Good agreement between $\overline {x_3}$ and the theoretical 
result for $x_3$ obtained for the three-nucleon system~\cite{bms1}
favours the second option.
 Accordingly, I conclude that it is essentially the binding
forces of the three-nucleon system which define the pattern
of the $F_2(x)$ modifications if $A >$ 4.

Taking into consideration a good agreement
between the theory and $^4$He/D data one can conclude that the overall
picture is consistent with the two stage evolution of the nucleon
structure as a function of $A$, one for $A \leq$ 4, 
and another one for $A >$ 4. Such a conception 
naturally explains experimentally observed factorization
of the $x$ and $A$ dependence on the $F_2(x)$ modifications.
It follows then that the partonic structure found for
the three-nucleon system can be understood
as the basic structure for all nuclear systems
with two exceptions: 1) deuteron, as a loosely bound system and
2) $^4$He, as an anomalously strongly bound system.
When the phenomenon is confirmed with experiments on $^3$He target
the EMC effect might obtain new non-trivial formulation: 
the pattern of  partonic structure, which is typical for 
metals, is identical to that of $^3$He and $^3$H.\\[-0.2cm]

The possibility of the factorization of the $x$ and $A$ dependence
in a restricted kinematic range 
has been discussed in early publications on the EMC effect. In
Ref.~\cite{jaffe84} the $A$ dependence of $r^A(x=0.55)$ has been
predicted by considering realistic nucleon density function obtained
from charge-density functions. The factorization has also been considered
in a number of subsequent publications~\cite{rep160} which
studied the effect in the range 0.3 $< x <$ 0.7 but did not relate it 
with the saturation of the binding forces in the lightest nuclei.

The factorization in the nuclear shadowing region has been introduced
in Ref.~\cite{wang} as an empirical relation to describe $r^A(x)$. 
A reasonable description of the data has required nine free parameters.

A somewhat different motivation of the factorization as compared to
the present paper and my previous analysis~\cite{sm95,sm94}, but still 
relevant either to the structure of the three-nucleon system or to the
property of nuclear binding, can be found in Refs.~\cite{bps,sick,bars85}.
The model of the three-gluon self-interaction in a three-nucleon system
has been suggested in Ref.~\cite{bars85}. It explains the factorization
in question and provides a physical basis for quantitative 
description of the data in the range 0.02 $< x <$ 0.65
when a five-parameter fit for the $r^A(x)$ is used~\cite{bars90}.
  
Very close to our conception of the two stage evolution is the
suggestion of Ref.~\cite{bps} to study the nucleon 
structure  modifications  in the infinite nuclear matter (INM).
The nuclear matter cross section is found from the finite-nucleus data
by extrapolating them to mass number $A = \infty$ using the $A^{-1/3}$
law. Such an approach strongly advocates for the nuclear binding
mechanism for describing the origin of the effect and 
has to rely on theoretical description of the INM. Quantitative
description of the effect can be smeared by the uncertainties of
the extrapolation which is not needed in our approach.

Thus, apart from this analysis and the theoretical consideration
of the  binding effects in the lightest nuclei~\cite{bms1}
it is only Refs.~\cite{bars85,bars90} which recognize,
however with different arguments,
a three-nucleon system as a decisive object for the understanding
of the EMC effect origin.\\[-0.3cm]

The results of the present paper may serve not only for the 
clarification of the role of the nuclear environment in the nucleon
structure, but also for a better understanding of the free nucleon
structure. Indeed, in the framework of the theory of $F_2(x)$
evolution~\cite{bms1}, $F_2^A(x)$ = $F_2^{\rm D}(x)$
if the sum of terms with $d F_2^{\rm D}(x)/dx$
and  $d F_2^{\rm N}(x)/dx$ changes sign.
Thus the positions of $x_i$ indicate the kinematic regions
in which it is desirable to increase accuracy of the $F_2(x)$ and to
measure it in fine $x$ binning. Evidently, one should think about
planning new  DIS experiments on proton, deuteron and $^3$He targets. 
Besides, the information on the derivatives of $F_2(x)$
might be used for realistic parametrization of the structure functions.
Similar considerations apply to the spin-dependent structure functions
$g_1^{\rm p}$ and $g_1^{\rm D}$. The spin degrees of freedom are
expected to magnify effects in the vicinity of $x_i$ 
due to the Pauli exclusion principle.

A plausible explanation of the correlations expressed with Eqs.~(15) and
(16) can be given by assuming a  decomposition of the obtained values
into contributions from nucleonic and partonic mechanisms.
One might suggest that the three-nucleon field produces a fairly small
redistribution of the partonic momenta in a bound nucleon in the
momentum range $x <$ 0.3, where  exchange pions are expected to
contribute to the nuclear binding forces. If one assumes that the
redistribution is of the order $m_{\pi}/3{\rm M}$
one finds that what is obtained from the present analysis 
can be reasonably well approximated as follows:
\begin{eqnarray}
1 - \overline {x_1}&&\approx~~1 ~ ~ ~ ~-~ m_{\pi}/3{\rm M},\\
1 - \overline {x_2}&&\approx~~ 2/3 ~+~ m_{\pi}/3{\rm M},\\ 
1 - \overline {x_3}&&\approx~~ 1/6 .
\label{x3m}
\end{eqnarray}
  
Position of $x_1$, as has been shown in Ref.~\cite{sm95},
is consistent with explanations of nuclear shadowing by an
overlap of partons belonging to a three-nucleon system.

Contrary to the situation with $x_1$ and $x_3$, the problem
of precise evaluation  of the second cross-over, $x_2$,
represents a challenge for theories and even for models of EMC effect.
Evidently, a parton redistribution effect represented by Eq.~(19)
%
is not a task to be solved either in a quark model or in
a conventional nuclear structure model alone.\\

{\bf 5~ Conclusions}\\[-0.2cm]

The world data on the EMC effect in the range of $A >$ 4 have been 
analyzed to determine the {\em pattern} of modifications of the
free nucleon structure function $F_2(x)$ in the nuclear environment.
It is found that the pattern is defined with the three $A$ independent
cross-over points.

I have obtained experimental evidence of the factorization
of the $x$ and $A$ dependence of the $F_2(x)$ modifications
for nuclei with $A >$ 4 in the entire range of $x$ which
signifies that distortions of parton distributions in nuclear
environment are saturated at $A \leq$  4. The phenomenon of saturation
is a natural consequence of the nuclear binding effects in $F_2(x)$,
which have been evaluated in a relativistic field theory of
nuclei ($A \leq$ 4) by Burov, Molochkov and Smirnov. Excellent
agreement with the available  $^4$He/D data allows one to conclude
that nuclear binding is the only physical mechanism responsible
for the EMC effect.

The agreement with the theory is even more spectacular when
predictions are confronted with $A \geq$ 4 data,
by simply scaling the modifications of $F_2(x)$ for $A$ = 3
with the $x$ independent factor defined by conventional nuclear
structure considerations. One can identify 
the partonic structure in the three-nucleon system found
by Burov, Molochkov and Smirnov as the basic structure for all
nuclear systems with two exceptions only: D and $^4$He.

The observation provides a clear-cut explanation of
the EMC effect origin: the nucleon partonic structure is modified
by {\em nuclear binding} forces and modifications are the strongest
in $^4$He. The partonic structure, which develops 
in {\em a three-nucleon system}, evolves to higher nuclear masses by
changing the amplitude of deviations of $F_2^A(x)/F_2^{\rm D}(x)$
from unity in full agreement with the variation of nuclear density
and geometry of a nucleus.\\

{\large\em Acknowledgements.} I would like to thank V.V.~Burov and
A.V.~Molochkov for helpful discussions of the problem of the 
relativistic binding effects in the structure of the lightest nuclei.
I have also benefited from discussions with S.V.~Akulinichev, A.M.~Baldin,
 K.~Eskola, H.G.~Fischer, S.B.~Gerasimov, R.L.~Jaffe,
S.A.~Kulagin and Yu.M.~Shabelskii. 
It is my pleasure to thank S.~Barshay, M.~Ericson and V.R.~Pandharipande
for some stimulating remarks.

\end{document}